\documentclass[sigconf]{acmart}

\usepackage{balance}

\usepackage[utf8]{inputenc} 
\usepackage[T1]{fontenc}    
\usepackage{hyperref}       
\usepackage{url}            
\usepackage{booktabs}       
\usepackage{amsfonts}       
\usepackage{nicefrac}       
\usepackage{microtype}      

\usepackage{amsmath}
\usepackage{amssymb}
\usepackage{amsthm}
\usepackage{comment}
\usepackage{enumerate}
\usepackage{algorithmic}
\usepackage{algorithm}
\usepackage{graphicx}
\usepackage{mathtools}
\usepackage{subfigure}
\usepackage{multirow}
\usepackage{tikz}
\usepackage{color}
\usetikzlibrary{positioning,shapes.geometric,arrows}

\newcommand*{\vimage}[1]{\vcenter{\hbox{\includegraphics[width=0.4\textwidth]{#1}}}}

\def\BibTeX{{\rm B\kern-.05em{\sc i\kern-.025em b}\kern-.08emT\kern-.1667em\lower.7ex\hbox{E}\kern-.125emX}}
    
\copyrightyear{2019} 
\acmYear{2019} 
\setcopyright{rightsretained} 
\acmConference[SIGIR '19]{Proceedings of the 42nd International ACM SIGIR Conference on Research and Development in Information Retrieval}{July 21--25, 2019}{Paris, France}
\acmBooktitle{Proceedings of the 42nd International ACM SIGIR Conference on Research and Development in Information Retrieval (SIGIR '19), July 21--25, 2019, Paris, France}
\acmDOI{10.1145/3331184.3331204}
\acmISBN{978-1-4503-6172-9/19/07}

\acmSubmissionID{fp055}

\begin{document}

\tikzstyle{rect} = [draw, rectangle, fill=white!20,text width=6em, text centered, minimum height=2em]
\tikzstyle{circ} = [draw, circle, minimum width = 12pt, inner sep = 17pt]

\title{Domain Adaptation for Enterprise Email Search}

\author{%
Brandon Tran$^{1*}$, Maryam Karimzadehgan$^2$, Rama Kumar Pasumarthi$^2$, Michael Bendersky$^2$, Donald Metzler$^2$
}
\affiliation{%
  \institution{$^1$Department of Mathematics, Massachusetts Institute of Technology, MA, USA}
}
\affiliation{%
  \institution{$^2$Google LLC, Mountain View, CA, USA}
}
\affiliation{%
  \institution{$^1$btran115@mit.edu $\quad$ $^2$\{maryamk, ramakumar, bemike, metzler\}@google.com}
}

\thanks{*Work done while Brandon Tran was at Google.}

%
\renewcommand{\shortauthors}{Trovato and Tobin, et al.}

%
\begin{abstract}

\label{sec:abs}

In the enterprise email search setting, the same search engine often powers multiple enterprises from various industries: technology, education, manufacturing, etc. However, using the same global ranking model across different enterprises may result in suboptimal search quality, due to the corpora differences and distinct information needs. On the other hand, training an individual ranking model for each enterprise may be infeasible, especially for smaller institutions with limited data. To address this data challenge, in this paper we propose a domain adaptation approach that fine-tunes the global model to each individual enterprise. In particular, we propose a novel application of the Maximum Mean Discrepancy (MMD) approach to information retrieval, which attempts to bridge the gap between the global data distribution and the data distribution for a given individual enterprise. We conduct a comprehensive set of experiments on a large-scale email search engine, and demonstrate that the MMD approach consistently improves the search quality for multiple individual domains, both in comparison to the global ranking model, as well as several competitive domain adaptation baselines including adversarial learning methods.

\end{abstract}

%
%

%

%
\maketitle

{\fontsize{8pt}{8pt} \selectfont
\textbf{ACM Reference Format:}\\
Brandon Tran, Maryam Karimzadehgan, Rama Kumar Pasumarthi, Michael Bendersky, Donald Metzler. 2019. Domain Adaptation for Enterprise
Email Search. In \textit{Proceedings of the 42nd Int'l ACM SIGIR Conference on Research and Development in Information Retrieval (SIGIR’19), July 21--25, 2019, Paris, France.} ACM, NY, NY, USA, 10 pages.https://doi.org/10.1145/3331184.3331204}

\section{Introduction}
\label{sec:intro}



\begin{figure}[!htp]
\begin{center}
\includegraphics[width=0.2\textwidth]{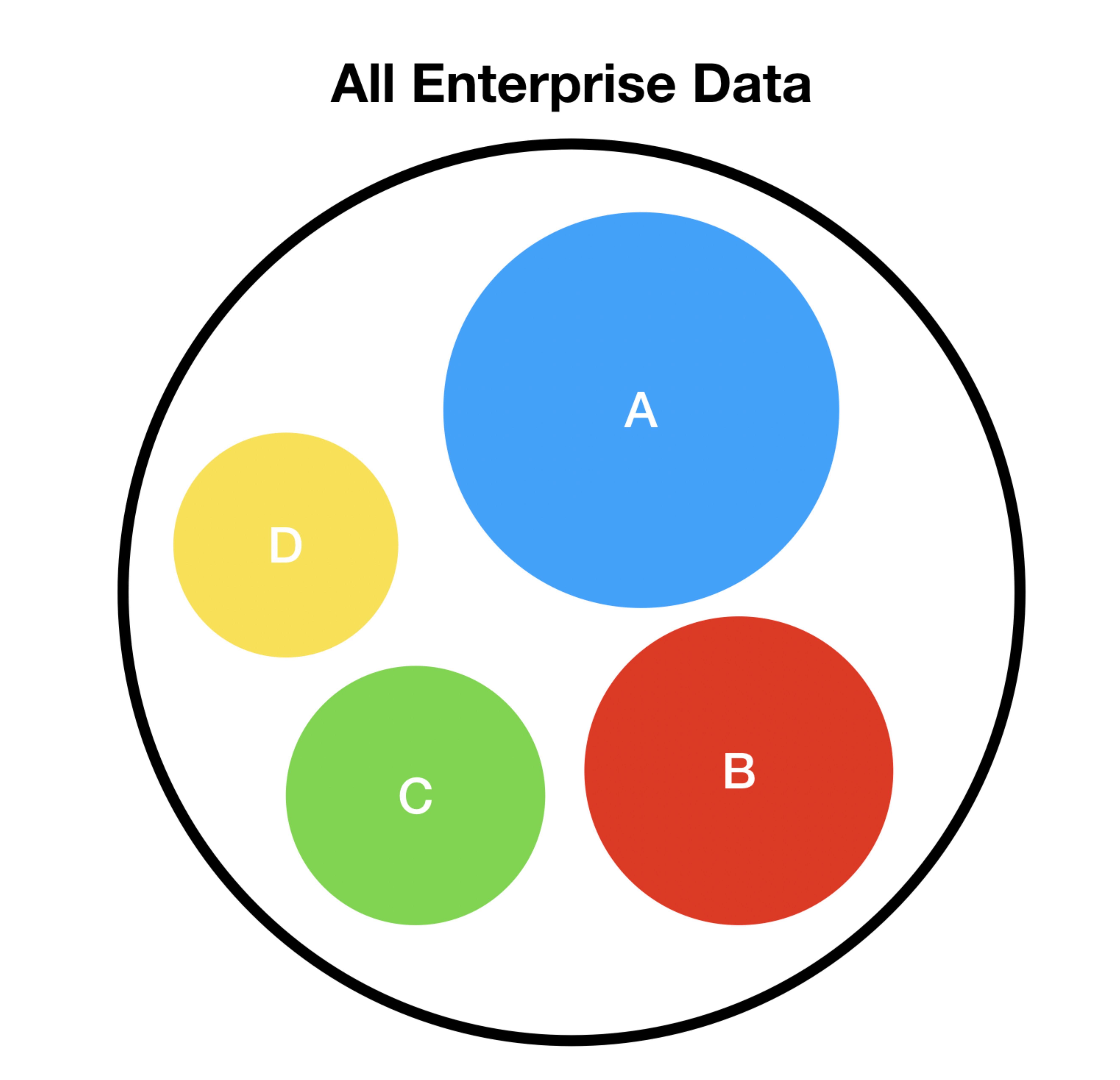}
\end{center}
\caption{A visualization of our enterprise email search data. The set of all data (the \emph{source domain}) consists of multiple individual enterprise domains (\emph{target domains}), with four examples in the figure labeled with A, B, C, and D.}
\label{fig:data}
\end{figure}

Traditionally, enterprise email search engines were installed locally on the premises of the organization. In these installations, search ranking was generally predetermined by the vendor and kept fixed. In the last several years, cloud-based search engines (e.g., Microsoft Azure, Amazon CloudSearch, or Google Cloud Search) have been gaining traction as an effective tool for search. In these cloud solutions, the corpora, the ranking models, and the search logs are stored in the cloud. This enables the cloud search providers to optimize the quality of their search engines based on user click data, similarly to what was previously done in web search~\cite{joachims02}. 

The transition to the cloud and the abundance of available user interaction data provide a unique opportunity to significantly improve the quality of enterprise email search engines, which traditionally lagged behind web search~\cite{Hawking:2010}. In particular, in recent years, deep neural learning-to-rank models were shown to significantly improve the performance of search engines in the presence of large-scale query logs, both in web search~\cite{guo16} and in email search~\cite{meltzer18,zamani17} settings. 

However, directly applying these advances in neural learning-to-rank models to enterprise email search is not straightforward. An important difference between web and enterprise email search is that in the latter, the model can be applied to multiple, often very different domains, as described in Figure~\ref{fig:data}. For instance, the same enterprise email search engine can power enterprises from various industries: technology, education, manufacturing, etc. Therefore, using the same global learning-to-rank model across the different enterprises may lead to suboptimal search quality for each individual enterprise, due to the differences between their corpora and information needs. 

On the other hand, training an individual ranking model for each industry or enterprise may not be feasible, especially for smaller enterprises with limited search traffic data. This is especially true for deep learning-to-rank models that typically require large amounts of training data. The reason for this requirement is that deep neural networks are susceptible to overfitting to the training data, and they need enough inputs to actually learn a model rather than simply memorizing the input examples.

As such, a natural question to ask is how to make use of the state-of-the-art deep learning-to-rank models in the enterprise search setting. To this end, we propose the use of domain adaptation techniques~\cite{ganin14,long15} to adapt a global model, trained on the entire data, to a particular enterprise domain. In this work, we specifically focus on the enterprise email search setting; however our findings easily generalize to other enterprise search scenarios as well.

Domain adaptation, at a high level, deals with the ability to adapt efficient, high-performing models trained on one domain to perform well on a different domain.
Typically, the first domain, called the source domain (in our case the entire dataset -- see Figure~\ref{fig:data}), contains a wealth of data, and so an effective prediction model can be trained.
However, due to what is known as dataset bias or domain shift~\cite{gretton09}, these models do not immediately generalize to new datasets, referred to as the target domains (in our case individual enterprises).
The target domains are expected to be significantly smaller than the source domain, so that a model cannot simply be trained with only target training data due to overfitting.
As a result, these fully-trained networks are typically fine-tuned to the new dataset.
That is, the available labeled data from the target domain is used to slightly alter the parameters of the original model to fit new data.
This is, of course, difficult and expensive to carry out.

Work in domain adaptation, then, attempts to reduce the harmful effects of this domain shift.
The deep learning model maps both the source and target domain into some latent feature space.
Reduction of the domain shift is then accomplished by either minimizing some measure of domain shift, such as maximum mean discrepancy (MMD)~\cite{tzeng14, long15}, or with adversarial adaptation methods~\cite{ganin14, tzeng15, liu16, tzeng17}.
In the latter case, the model is trained to make the two mappings of the source and target domain indistinguishable in feature space to a discriminator model.

While domain adaptation methods have been previously studied, we emphasize that the majority of the research done was in other areas such as image classification.
Moreover, the typical focus was on an unsupervised setting with no labeled data from the target domain. We note, though, that there has been one related work by Cohen et al.~\cite{cohen18}, studying the problem of domain adaptation in learning-to-rank models.

However, there are two major differences between these prior works, including~\cite{cohen18}, and the problem we study.
First, the enterprise email search problem deals with datasets that are \emph{several orders of magnitude} larger than the previous work.
The source domain, consisting of the combined inputs from all individual enterprise domains, contains $O(100M)$ inputs as compared to $O(10K)$ inputs in the image classification datasets~\cite{tzeng14, long15, ganin14, tzeng15, liu16, tzeng17} and the prior learning-to-rank work~\cite{cohen18}. 
Additionally, we deal with labeled enterprise domains in a weakly supervised setting (using user click data), whereas the aforementioned prior works all assumed unlabeled target domains.

These differences from prior work lead to a more realistic setting for exploring domain adaptation techniques in information retrieval. As we show through extensive experimentation, in this setting MMD outperforms all other domain adaptation techniques, including the state-of-the-art adversarial methods~\cite{cohen18}, a first such result in the information retrieval literature. In summary, our key contributions are:

\begin{itemize}
    \item We propose a general framework for learning-to-rank with domain adaptation, with a particular application to enterprise email search.
    \item We experimentally demonstrate the shortcomings of simple transfer learning methods, such as re-training or batch-balancing, to individual enterprise domains.
    \item We propose a novel use of the maximum mean discrepancy (MMD) method for learning-to-rank with domain adaptation, and demonstrate its effectiveness and robustness in the enterprise email search setting.
    \item We perform a thorough comparative analysis of various domain adaptation methods on realistic, large-scale enterprise email search data.
\end{itemize}


The rest of our paper is organized as follows.
First, we discuss related work in Section~\ref{sec:rel}.
Next, in Section~\ref{sec:mot}, we provide motivating evidence that using domain adaptation techniques is feasible for our setting.
That is, we show that the distributions of representations of the source and target datasets have nontrivial overlap, and so it is reasonable to try and encourage the model to accurately predict clicks on both sets.
In Section~\ref{sec:methodolgy}, we give a detailed explanation of our methodology.
Then, we provide our extensive experimental study in Section~\ref{sec:exp}.
Finally, we conclude and discuss future work in Section~\ref{sec:conc}.


\section{Related Work}
\label{sec:rel}
We split our discussion of related works into four distinct parts.
First, we discuss work done in the learning-to-rank literature on which we base our learning-to-rank models. Next, we review research done in enterprise email search setting. 
Then, we mention work done in developing techniques for domain adaptation in image classification.
Finally, we point out other works trying to utilize domain adaptation for information retrieval problems.

\subsection{Learning-to-Rank} 
Generally, learning-to-rank refers to the application of machine learning tools and algorithms to rank models for information retrieval.
There is a vast literature of learning-to-rank work~\cite{burges05, burges10, cao07, friedman01, joachims02, xia08}, differing in their model and loss function constructions.
Recently, with the rise in popularity of deep neural networks, work has been done to use deep neural networks for learning-to-rank~\cite{dehghani17, borisov16}. For a complete literature review on neural ranking models for information retrieval, please refer to the survey by Mitra~\cite{Mitra2017}.

\subsection{Enterprise Search}
Enterprise search can broadly be viewed as the application of information retrieval techniques towards the specific problem of searching within private organizations.
The majority of prior work can be found in the recent surveys by Kruschwitz and Hull~\cite{Kruschwitz+Hull:2017} and Hawking~\cite{Hawking:2010}.
To the best of our knowledge, no previous research on enterprise search has studied the problem as an application of domain adaptation.

Enterprise search is also closely related to personal search (e.g., email search), as both deal with searching in private or access controlled corpora~\cite{ai17, carmel15, dumais16, grevet14, wang16, meltzer18, kim17}.
Even though some success has been found using time-based approaches for personal search~\cite{dumais16}, relevance-based ranking arising from learning-to-rank deep neural network models has become increasingly popular~\cite{zamani17,meltzer18} as the sizes of private corpora increase~\cite{grevet14}. However, to the best of our knowledge, our work is the first study on applying deep neural networks specifically in the enterprise search setting.

\subsection{Domain Adaptation} 
Extensive prior work on domain adaptation has been done in image classification~\cite{gretton09}.
These works develop methods to transfer latent representations obtained from deep neural networks from a large, labeled source dataset to a smaller, unlabeled target dataset.
The primary strategy focuses on guiding the learning process by encouraging the source and target feature representations to be indistinguishable in some way~\cite{tzeng14, long15, sun16, ghifary16, ganin14, tzeng15, liu16}.

One focus of domain adaptation research in image classification aims on minimizing the differences of certain statistics between the source and target distributions.
Several methods used the Maximum Mean Discrepancy (MMD)~\cite{gretton09} loss, in a variety of ways~\cite{tzeng14, long15, sun16}.
The MMD computes the norm of the difference between two domain means.
Another direction lies in choosing adversarial losses to minimize domain shift.
The goal is to learn a representation from which a classifier can predict source dataset labels while a discriminator cannot distinguish whether a data point is from the source or target datasets.
The work in this direction focuses on finding better adversarial losses~\cite{ghifary16, ganin14, tzeng15}.

\subsection{Applications of Domain Adaptation to Information Retrieval}
\label{sec:rel_dair}
As deep learning models became more prevalent in ranking problems, and as more and more transfer learning techniques developed for image classification, work began to study the problem of domain adaptation in the information retrieval domain.
The models in~\citet{cohen18} are trained for domain adaptation in ranking with adversarial learning.
Specifically, the models are trained using the gradient reversal layers of~\citet{ganin16}.
We note that this work focused only on adversarial learning and did not consider maximum mean discrepancy.
Moreover, it only compared their adversarial learning technique with very simple baselines: training on all data and training on target data.
The work did not consider more interesting baselines such as balancing training batches with a certain number of target data inputs, or fine-tuning a previously trained model on all data with only target data. 
Additionally, while in the learning-to-rank setting, their datasets are significantly smaller and not as complex as those arising from enterprise search.
Similarly, the work of~\citet{long18} also utilizes adversarial learning to solve the problem of domain adaptation to a number of information retrieval tasks, including digit retrieval and image transfers.
However, as with~\citet{cohen18}, they study only adversarial learning, and their data are also significantly different from enterprise search.
Lastly,~\citet{mao18} use a similar statistics-based approach for transferring features from source to target domains.
Their method also looks at a certain mean distance computed from the embedded data, but their per-category multi-layer joint kernelized mean distances are quite distinct from an MMD regularization term.


\section{Motivation}
\label{sec:mot}

In this section, we provide motivation for using domain adaptation techniques in enterprise email search.
First, we take our enterprise email search inputs and map them into a high-dimensional space. 
As detailed in Section~\ref{sec:embed}, we refer to the resulting subset of the high-dimensional space obtained from this mapping as the embedding of our inputs.
These embeddings are then passed as inputs into the prediction models, as discussed in Section~\ref{sec:dnn}.
The high-level goal of domain adaptation techniques is to make the embeddings arising from the source and target domains indistinguishable to the prediction models.
That way, the model can leverage information from the source domain in order to make predictions on the target domain.
However, these techniques crucially rely on the embeddings for the source and target datasets to take a certain form.
First, the embeddings cannot be distributionally identical.
If this were the case, simply training on all the data or even just the source data would yield a model that generalizes to the target data.
Second, the embeddings must have nontrivial overlap.
The model can, via gradient descent updates to the embedding weights, push the two distributions closer together.
But if they are too far apart to begin with, one cannot hope for this to be successful.

To this end, we present some experimental results to show that our data does indeed take the required form.
We take a network trained on all data to completion and study the embedding distributions of the source and target datasets.
To reiterate what we said in Section~\ref{sec:intro}, the source dataset here refers to the entire search log data, and the target dataset is a specific small enterprise domain.
First, we compute the means of the two distributions and compare their norms~\cite{mohri18} to the norm of their difference.
From Table~\ref{tab:norms}, we can see that the two mean vectors and the difference vector all have a norm that is of the same order.
This suggests that the means of the two distributions are indeed quite different, providing evidence that indeed, the source and target domain embeddings are not distributionally identical.

\begin{table}[!ht]
\caption{Table of norms of the source and target dataset embedding means as well as their difference.}
\label{tab:norms}
\begin{center}
{\renewcommand{\arraystretch}{1.4}
\begin{tabular}{|cc|}
  \hline
  Source Mean Norm:& 1.0578 \\
  \hline    
  Target Mean Norm:& 1.3558 \\
  \hline    
  Norm of Mean Difference:& 0.8558\\
  \hline
\end{tabular}}
\end{center}
\end{table}

\begin{figure}[!htp]
\begin{center}
\resizebox{.8\linewidth}{.4\linewidth}{
\includegraphics{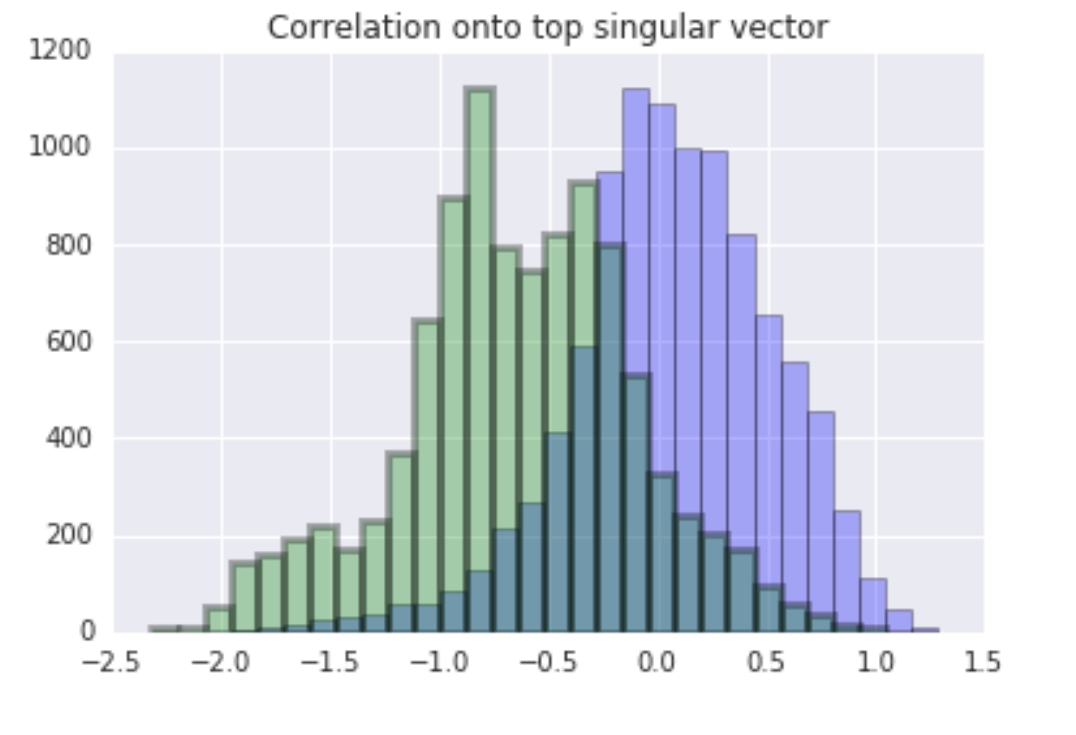}
}
\end{center}
\caption{Some statistics of the two embedding distributions from source and target datasets. There are $10,000$ examples living in a $508$-dimensional space. We show the distributions projected onto the direction of the largest eigenvector of the covariance matrix of embedding vectors. The source dataset is in blue, and the target dataset is in green.}
\label{fig:mot}
\end{figure}

Additionally, while we cannot visualize the distributions in multidimensional space, we can plot their projections onto a one-dimensional space.
To choose the vector on which we project the distributions, we use an idea from robust statistics~\cite{diakonikolas16, lai16}.
We form a matrix consisting of the embedding vector of each example from both the source and target datasets.
From this matrix, we compute the vector corresponding to the largest singular value.
The intuition is that this vector corresponds to the strongest signal coming from the combined embedding vectors. 
If the source and target embeddings indeed form two distinct populations, this signal will correlate more strongly with one distribution than the other.
From Figure~\ref{fig:mot}, we can see that this is the case.
The green values represent the correlations of the target set embeddings onto the top singular vector, and the blue values are from the source data embeddings.
While there is some overlap between the two distributions, they are quite clearly distinct.

As we have now established that the embedding distributions are overlapping, but not identical, in the next section we discuss possible domain adaptation techniques to be used.


\section{Methodology}
\label{sec:methodolgy}

In this section, we first formulate our problem and provide definitions of notation we will use. Then, we describe two different solutions to the problem, first using discriminator-based techniques and then statistics-based techniques.

\subsection{Problem Formulation}
The inputs for the enterprise email search problem are queries $\psi=(q,D,c)$, where $q$ represents a user query string, $D$ represents the query's list of documents, and $c$ represents the clickthrough data.
For each document $d\in D$, we have a feature vector $x_d$ as its representation, and a boolean $c_d$ denoting whether the document was clicked.
The set of queries from all data is labeled $\mathcal{S}$, the source dataset, while the set of queries from a specific domain is labeled $\mathcal{T}$, the target dataset, and we are trying to use $\mathcal{S}$ to help train a model on $\mathcal{T}$.

Given the set of queries, our goal is to learn a model $M$ minimizing a loss defined as:
\begin{equation}
\mathcal{L}(M) = \mathrm{E}_{\psi\in\mathcal{T}} (\ell(M,\psi)), \label{eq:true_loss}
\end{equation}
where $\ell(M,\psi)$ denotes the loss of the model $M$ on query $\psi$, and we are taking expectation over queries from the target data.
Before we define our loss function, we note that typical neural network models work by approximating the above loss with the training set.
We establish notation here by letting the training sets for our deep network be $S=\{\psi_i^S\} \in\mathcal{S}$ and $T=\{\psi_j^T\} \in\mathcal{T}$. 
Our goal is to learn a model $M$ that minimizes:
\begin{equation*}
\mathcal{L}(M) = \frac{1}{|T|} \sum_{\psi^T\in T} \ell(M,\psi^T).
\end{equation*}

However, in the domain adaptation setting, we assume a scarcity of data in the target distribution.
As such, training $M$ to minimize $\mathcal{L}$ would result in either an overfitted model or one that cannot generalize to all of $\mathcal{T}$.
Thus, we instead train $M$ using training data from both $\mathcal{S}$ and $\mathcal{T}$.

In this paper, our model $M$ depends on deep neural networks (DNNs)~\cite{lecun15}.
We choose to use DNNs for a few reasons.
First, the number of features from our queries and documents is quite large, and moreover, some features are sparse.
While tree-based models~\cite{friedman01} can model feature interactions, they are not scalable to a large number of features and cannot handle sparse features, such as those coming from document or query text.
On the other hand, DNNs are a natural candidate model for dealing with sparse inputs.
Also, DNNs were shown to be quite successful for ranking applications, especially when supplied with large amounts of supervised training data~\cite{dehghani17}.

Over the next few sections, we provide an overview of our model $M$.
First, we map the query and document features together into a high-dimensional embedding space (specifically, $508$-dimensional).
Then, a prediction model, which we will call $P$, consisting of a DNN is trained on this embedding space.
Since the source and target datasets are different, we also expect their embeddings to be different within the embedding space.
Thus, we use an additional correction model to make the embeddings indistinguishable to the prediction model, so that source data can be used to predict clicks on the target data.

\begin{figure*}[!htp]
\begin{center}
    \scalebox{.8}{
    \begin{tikzpicture}[node distance = 1.5cm, auto]
    \node [] (top) {};
    \node [rect, rounded corners, fill=blue!20!white, below=0.3cm of top] (source) {Source Dataset};
    \node [rect, rounded corners, fill=red!20!white, below=0.8cm of source] (target) {Target Dataset};
    \node [below=0.4cm of source] (mid) {};
    \node [rect, rounded corners, minimum height=10em, right=3cm of mid, label=above:Embedding Space, label=below:$(1)$] (embed) {};
    \begin{scope}[blend group=multiply]
        \node [circle, fill=blue!20!white, minimum size=1.8cm] at (4.2,-1.2) (sembed) {};
        \node [circle, fill=red!20!white, minimum size=1.8cm] at (4.2,-2.2) (tembed) {};
    \end{scope}
    \node [rect, rounded corners, right=6.5cm of source, label=below:$(2)$] (ranker) {Prediction Model (P)};
    \node [rect, rounded corners, right=6.5cm of target, label=below:$(3)$] (correction) {Correction Model};
    \node [rect, rounded corners, right=6.5cm of embed, minimum height = 3em] (loss) {Training Loss};
    \path[->]
        (source) edge node[inner sep=2pt, align=left, sloped, above] {embed} (sembed);
    \path[->]
        (target) edge node[inner sep=2pt, align=left, sloped, above] {embed} (tembed);
    \path[->]
        (embed) edge node[inner sep=2pt, align=left, sloped, above] {Neural Net} (ranker);
    \path[->]
        (embed) edge node[inner sep=2pt, align=left, sloped, above] {} (correction);
    \path[->]
        (ranker) edge node[inner sep=2pt, align=left, sloped, above] {Softmax} (loss);
    \path[->]
        (correction) edge node[inner sep=2pt, align=left, sloped, above] {Regularizing} (loss);
    \path[->]
        (correction) edge node[inner sep=2pt, align=left, sloped, below] {Loss} (loss);
    \node[above,font=\large\bfseries] at (current bounding box.north) {Enterprise Search Model (M)};
    \end{tikzpicture}
    }
    \caption{Illustration of the Domain Adaptation Problem. Both the source and target datasets are mapped into a high-dimensional embedding space. Given the embeddings, a prediction model, i.e. a deep neural net, is trained to predict clicks using a softmax cross entropy loss. For domain adaptation, a correction model is used on the embedding space to compute a regularizing loss term that is added to the training loss.}
    \label{fig:diagram}
\end{center}
\end{figure*}
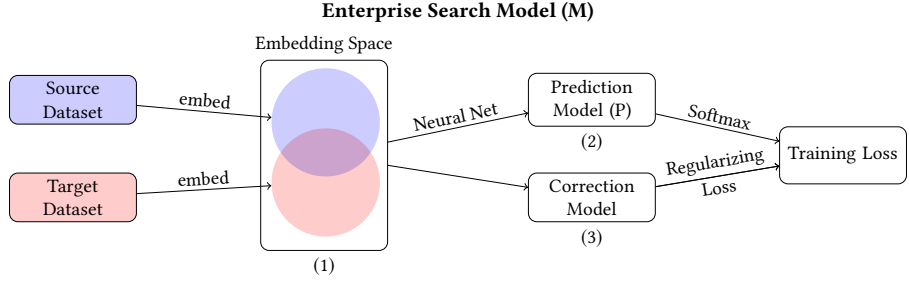

Figure~\ref{fig:diagram} provides an illustration of our problem.
In the next section, we describe how to map enterprise email search inputs into a high-dimensional embedding space.

\subsection{Input Embeddings}
\label{sec:embed}
The source and target datasets are mapped to an embedding space via an embedding function (part $(1)$ in Figure~\ref{fig:diagram}).

Due to the private nature of our data, our query-document inputs are modeled as bags of frequent n-grams and can be represented in three parts: 
\begin{itemize}
    \item the features for the query $q$
    \item the sparse features for a document $d$
    \item the dense features for $d$ (e.g., document static scores). 
\end{itemize}

To preserve privacy, the inputs are k-anonymized and only query and document n-grams that are frequent in the entire corpus are retained. For more details about the specific features and the anonymization process used in a similar setting see~\cite{meltzer18,zamani17}.
The query features and sparse document features are passed through a single embedding layer, whereas the dense features are left as they are.
We denote the various parts of the input as $x_q^{\textrm{embed}}, x_d^{\textrm{embed}}$, and $x_d^{\textrm{dense}}$, concatenate them together, and then pass them through one additional embedding layer to form the input embedding, $x_d$.
We denote the function taking an input $\psi$ and outputting its corresponding embedding $x_d$ as $\textrm{emb}(\psi)$.

\subsection{Feed-Forward DNN Model}
\label{sec:dnn}

The discussion of our feed-forward DNN model corresponds to part $(2)$ of Figure~\ref{fig:diagram}, the prediction model, $P$.

Feed-forward DNNs have become increasingly prominent in research for learning-to-rank problems~\cite{edizel17, huang13, zamani17}.
Such models have proven to be effective in modeling sparse features naturally arising from inputs such as document text through the use of embeddings.
Since they are the basis for our models, we review them in this section. 

A deep neural network can be broken down into a series of layers, each applying a matrix multiplication, bias addition, and then activation function to its inputs.
We refer to the output of layer $i$ as $h_i$.
Letting $h_0$ be $x_d$, the subsequent layers are explicitly obtained as:

\begin{equation*}
h_i = \sigma(w_i h_{i-1} + b_i),
\end{equation*}
where $w_i$ denotes the weight matrix and $b_i$ the bias vectors at layer $i$.
We refer to the trainable parameters $w$ and $b$ together as the prediction model's $\theta_P$.
The function $\sigma$ is what is known as the activation function.
We use a hyperbolic function:
\begin{equation*}
    \sigma(t) = \dfrac{e^{2t}-1}{e^{2t}+1}.
\end{equation*}
If $h_{o}$ is the last layer, then the prediction model output is simply:
\begin{equation*}
    P(\textrm{emb}(\psi)) = P(x_d) = w_o h_o + b_o.
\end{equation*}

The loss function we aim to minimize utilizes a softmax cross entropy term. 
Formally, for an input $\psi$ with $N$ documents $d_i$ and click data $c_i$, the loss can be written as:
\begin{equation}
    \ell_P(\psi, \textrm{emb}) = - \sum_{i=1}^N c_{i}\log(p_i), \label{eq:ce_loss}
\end{equation}
where
\begin{equation*}
    p_i = \dfrac{1}{1+e^{-P(x_{d_i})}}.
\end{equation*}
The overall prediction model loss term is then the average of $\ell_P$ over all inputs $\psi$:
\begin{equation}
    \mathcal{L}_P(\{\psi\}, \textrm{emb}) = \frac{1}{|\{\psi\}|} \sum_{\psi} \ell_P(\psi, \textrm{emb})
    \label{eq:p_loss}
\end{equation}


\subsection{Domain Adaptation Methods}
\label{sec:methods}

In this section, we describe our techniques for encouraging the neural network models to make the embeddings of the source and target datasets indistinguishable. 
This corresponds to part $(3)$ of Figure~\ref{fig:diagram}, the correction model.
Our first class of methods in Section~\ref{sec:disc_tech} is motivated by Generative Adversarial Networks~\cite{goodfellow14}, relying on the deep neural networks called discriminators that arise in Generative Adversarial Networks.
Then, in Section~\ref{sec:stats_methods}, we propose a second class of methods focusing on utilizing various statistics of the embedding distributions for domain adaptation.

\subsubsection{Discriminator-Based Techniques}
\label{sec:disc_tech}

Discriminator-based domain adaptation techniques are closely related to adversarial learning methods where a two-player game is created between a discriminator and an adversary.
In this setting, each training example is labeled as being from either the source or the target dataset.
The discriminator is implemented as a neural network that works to classify an example with its corresponding dataset.
At the same time, an adversary updates the embedding weights in such a way as to fool the discriminator.

The goal of discriminator-based techniques is to reach an equilibrium in which the adversary has updated the embedding weights to fool any discriminator.
In this case, the two embeddings will be indistinguishable, and a prediction model trained on the source dataset will generalize well to the target dataset.

We then define the total loss function for discriminator-based techqniues on the model $M$ as follows:
\begin{equation}
    \mathcal{L} = \mathcal{L}_{P} + \lambda_D \mathcal{L}_D + \lambda_{adv} \mathcal{L}_{adv}.
\label{eq:grloss}
\end{equation}
Here, $\mathcal{L}_{P}$ refers to Equation~\ref{eq:p_loss}, while $\mathcal{L}_{D}$ and $\mathcal{L}_{adv}$ refer to the discriminator and adversarial losses which we will discuss next..
The $\lambda_D$ and $\lambda_{adv}$ terms are multiplicative factors that control the effects of the discriminator and adversarial losses relative to $\mathcal{L}_P$.

The discriminator itself is an additional feed-forward deep neural network, separate from the prediction model, taking the embedding of a query as input, which we will denote as a function $\mathcal{D}(\textrm{emb}(\psi))$.
Similar to the parameters for the prediction model, we will denote the trainable parameters for this DNN with $\theta_D$.
The discriminator loss (i.e., $\mathcal{L}_D$) is a standard cross-entropy loss, defined as:
\begin{align}
    \mathcal{L}_{D}(\{\psi_i^S\},\{\psi_j^T\},\textrm{emb}) = & -\frac{1}{|S|}\sum_{i=1}^{|S|} \log (D(\textrm{emb}(\psi_i^S))) \nonumber \\ 
    & - \frac{1}{|T|}\sum_{j=1}^{|T|} \log(1-D(\textrm{emb}(\psi_j^T))). \label{eq:disc}
\end{align}

For adversarial loss (i.e., $\mathcal{L}_{adv}$), there are different choices that can be made. 
One standard choice is known as the gradient reversal loss~\cite{ganin16}.
The idea with gradient reversal is to directly maximize the discriminator loss, i.e.,  $\mathcal{L}_D$. 
The gradient of the discriminator loss is, by definition, the direction of the largest change.
While the discriminator will take a gradient step to decrease the loss, the adversary takes a backwards step along this direction.
Formally, the adversarial loss is defined as follows:
\begin{equation}
    \mathcal{L}_{adv}(\{\psi_i^S\},\{\psi_j^T\},\textrm{emb}) = -\mathcal{L}_{D}(\{\psi_i^S\},\{\psi_j^T\},\textrm{emb}). \label{eq:adv}
\end{equation}

For completeness, we mention that there are two other often-used adversarial losses.
One uses a cross-entropy term, but with inverted labels~\cite{goodfellow14}, labeling each source example as coming from the target dataset, and vice versa.
The other computes the cross-entropy of the combined source and target datasets against a uniform distribution drawn from both~\cite{tzeng15}.
In our experiments, we found that these different losses all yielded similar performance and so focus on gradient reversal, which was also used successfully in a related work in domain adaptation for information retrieval~\cite{cohen18}. 

\begin{figure}[t!]
\begin{center}
\tikzstyle{vertrect} = [draw=black, rectangle, align=center, minimum width = 0.2cm, minimum height = 2.0cm]
\tikzstyle{myarrows}=[line width=0.5mm,draw=black,-triangle 45,postaction={draw, line width=2mm, shorten >=3mm, -}]
\scalebox{.9}{
    \begin{tikzpicture}[node distance = 1.5cm, auto]
    \node [fill=white, vertrect, rotate=-90, minimum height = 0.2cm, minimum width = 2.0cm] at (0,0) (input) {Input};
    \node [fill=green!20!white, vertrect] at (1,0) (em1) {};
    \node [fill=green!20!white, vertrect] at (2,0) (em2) {};
    \node [fill=green!20!white, vertrect, rotate=-90, minimum height = 0.2cm, minimum width = 2.0cm] at (3,0) (emb) {Embedding};
    \node [fill=red!20!white, vertrect] at (4,0) (fc1) {};
    \node [fill=red!20!white, vertrect] at (5,0) (fc2) {};
    \node [fill=red!20!white, vertrect] at (6,0) (fc3) {};
    \node [fill=red!20!white, vertrect, rotate=-90, minimum height = 0.2cm, minimum width = 2.0cm] at (7,0) (cl) {Click Labels};
    \node [fill=red!20!white, vertrect, rotate=-90, minimum height = 0.2cm, minimum width = 2.0cm] at (8,0) (clloss) {Loss $\mathcal{L}_P$};
    \node [fill=blue!20!white, vertrect] at (4,-2.2) (d1) {};
    \node [fill=blue!20!white, vertrect] at (5,-2.2) (d2) {};
    \node [fill=blue!20!white, vertrect] at (6,-2.2) (d3) {};
    \node [fill=blue!20!white, vertrect, rotate=-90, minimum height = 0.2cm, minimum width = 2.0cm] at (7,-2.2) (d) {Set Labels};
    \node [fill=blue!20!white, vertrect, rotate=-90, minimum height = 0.2cm, minimum width = 2.0cm] at (8,-2.2) (dloss) {Loss $\mathcal{L}_D$};
    \node [] at (0,1.1) (intop) {};
    \node [] at (3,1.1) (emtop) {};
    \node [] at (8,1.1) (cltop) {};
    \node [] at (0,-1.1) (inbot) {};
    \node [] at (3,-3.3) (embot) {};
    \node [] at (8,-3.3) (clbot) {};
    \draw[myarrows, draw=green!20!white] (input) -- (em1);
    \draw[myarrows, draw=green!20!white] (em1) -- (em2);
    \draw[myarrows, draw=green!20!white] (em2) -- (emb);
    \draw[myarrows, draw=red!20!white] (emb) -- (fc1);
    \draw[myarrows, draw=red!20!white] (fc1) -- (fc2);
    \draw[myarrows, draw=red!20!white] (fc2) -- (fc3);
    \draw[myarrows, draw=red!20!white] (fc3) -- (cl);
    \draw[myarrows, draw=blue!20!white] (emb) -- (d1);
    \draw[myarrows, draw=blue!20!white] (d1) -- (d2);
    \draw[myarrows, draw=blue!20!white] (d2) -- (d3);
    \draw[myarrows, draw=blue!20!white] (d3) -- (d);
    \path[->] (emtop) edge node[inner sep=2pt, align=left, above] {$\frac{\partial \mathcal{L}_P}{\partial \theta_{\textrm{emb}}}$} (intop);
    \path[->] (cltop) edge node[inner sep=2pt, align=left, above] {$\frac{\partial \mathcal{L}_P}{\partial \theta_P}$} (emtop);
    \path[->] (embot) edge node[inner sep=2pt, align=left, below] {$\frac{-\partial \mathcal{L}_D}{\partial \theta_{\textrm{emb}}}$} (inbot);
    \path[->] (clbot) edge node[inner sep=2pt, align=left, below] {$\frac{\partial \mathcal{L}_D}{\partial \theta_D}$} (embot);
    \end{tikzpicture}
    }
    \caption{Illustration of the Gradient Reversal Algorithm. We color code the overall model into three separate parts- green for the embedding, red for the prediction, and blue for the discriminator. An input is mapped by right-facing arrows to the prediction model and discriminator loss terms, $\mathcal{L}_P$ and $\mathcal{L}_D$ respectively. Then, during the backpropagation of gradients to train the neural network, each part of the network receives gradients as listed in the diagram. The discriminator part of the network is trained with gradients from $\mathcal{L}_D$, and the prediction model is trained with gradients from $\mathcal{L}_P$. The embedding part is trained with gradients from both.}
    \label{fig:disc}
\end{center}
\end{figure}
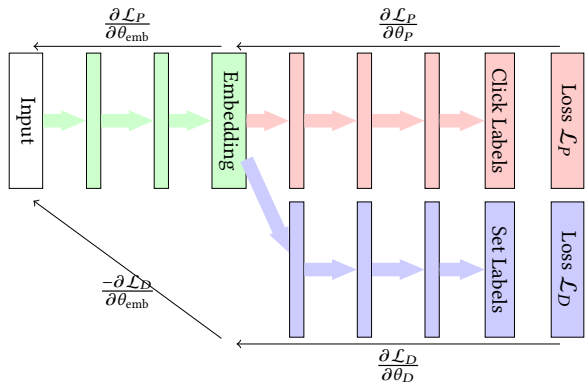

Since we will focus on gradient reversal, we provide an illustration of the technique in Figure~\ref{fig:disc} and will henceforth refer to our proposed training method from the class of discriminator-based techniques as the \textbf{gradient reversal method}.
The embedding parameters are $\theta_{\textrm{emb}}$ (green), the prediction model parameters are $\theta_P$ (red), and the discriminator parameters are $\theta_D$ (blue).
The gradient updates in each time step are given in the figure as a partial derivative of the loss functions with respect to the parameters.

\subsubsection{Statistics-Based Techniques}
\label{sec:stats_methods}

\begin{figure}[t!]
\begin{center}
{\renewcommand{\arraystretch}{1.4}
\begin{tabular}{c}
Embedding before MMD Minimization \\
\frame{\includegraphics[width=0.27\textwidth]{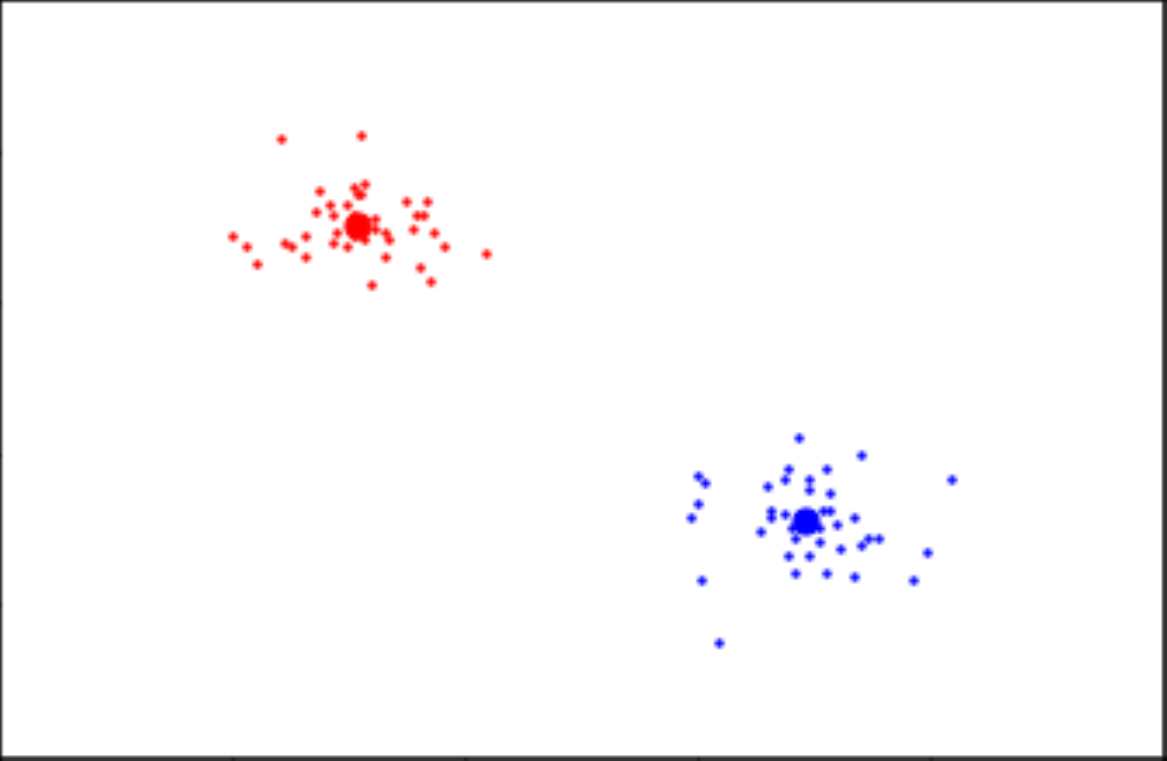}} \\
\resizebox{0.03\textwidth}{15pt}{$\Downarrow$}\\
Embedding after MMD Minimization \\
\frame{\includegraphics[width=0.27\textwidth]{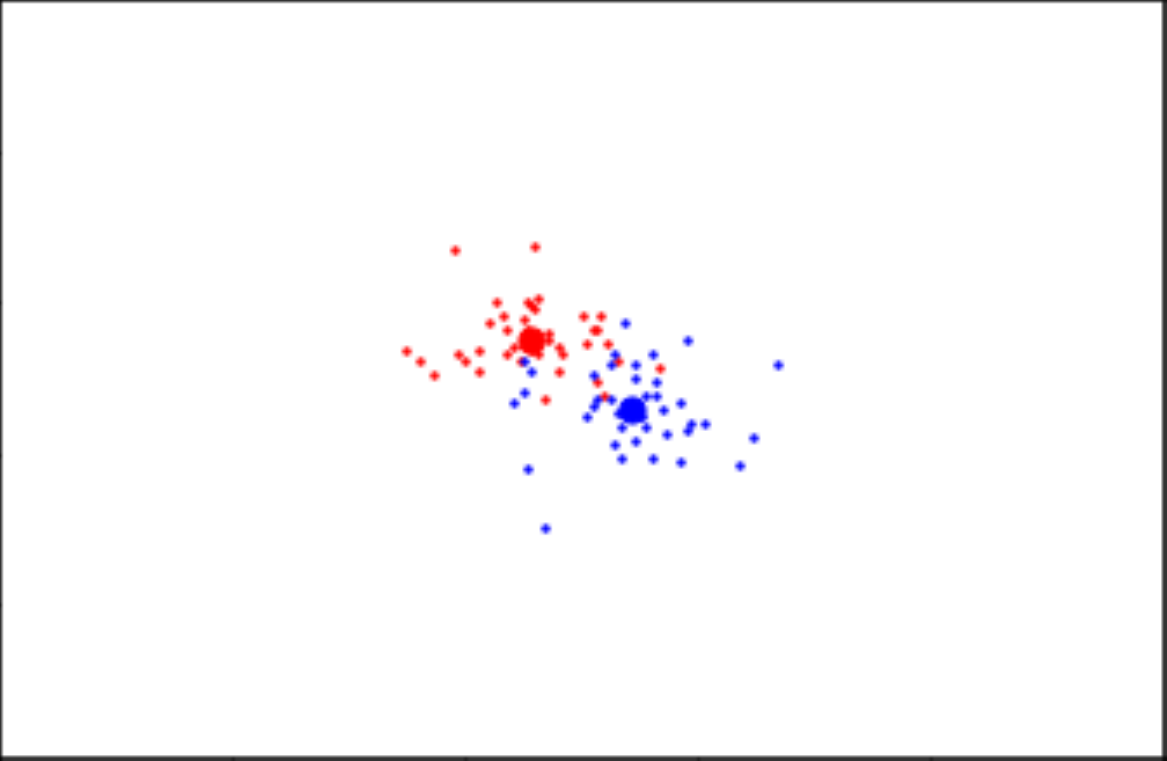}}\\
\end{tabular}}
\end{center}
\caption{Illustration of Maximum Mean Discrepancy Technique. The two distributions, labeled blue and red, are far apart in embedding space. The Maximum Mean Discrepancy technique adds a regularizing loss that encourages the model to minimize the distance between the means. Note that the distributions themselves do not change in shape. They are only brought closer together in embedding space.}
\label{fig:mmd}
\end{figure}

The embeddings from the source and target data points are subsets in our embedding space, which we can think of as distributions. 
Consequently, we can extract various statistics from the distributions and encourage the model to match the statistics coming from the two datasets.
We focus on using the mean of the distributions.
While two distributions can have the same or similar means but still be very different, we found empirically that this statistic actually worked quite well in making the two distributions indistinguishable to the prediction model.
Specifically, we add a term to the model loss function consisting of the difference in the means of the source and target embedding distributions.
Since neural networks work by minimizing their loss functions, this allows the network to take steps to minimize the difference in the means, drawing the two distributions closer together. 
We provide a pictorial representation of this technique in Figure~\ref{fig:mmd} which we refer to as \textbf{maximum mean discrepancy} (MMD).
The two distributions are given in red and blue, and their means are represented by a bold point.
Applying an MMD minimization does not change the shape of either distribution, but brings their means closer together.

The total loss for the model is then defined as:
\begin{equation}
    \mathcal{L} = \mathcal{L}_P + \lambda_{MMD} \mathcal{L}_{MMD}.
\label{eq:mmdloss}
\end{equation}
Where $\mathcal{L}_P$ is the prediction model loss (i.e., Equation~\ref{eq:p_loss}) and $\mathcal{L}_{MMD}$ is the maximum mean discrepancy loss. The $\lambda_{MMD}$ factor also controls the effect of the MMD on the overall loss relative to the prediction loss.

Formally, the MMD loss, i.e., $\mathcal{L}_{MMD}$ is given by:
\begin{align}
    \mathcal{L}_{MMD}(\{\psi_i^S\},\{\psi_j^T\},\textrm{emb}) = \Big|\Big|\frac{1}{|S|} \sum_{i=1}^{|S|} \textrm{emb}(\psi_i^S) 
    - \frac{1}{|T|} \sum_{j=1}^{|T|} \textrm{emb}(\psi_j^T)\Big|\Big|_2, \label{eq:mmd}
\end{align}

Since we are attempting to make two distributions indistinguishable in latent space, it is natural to also include other statistics.
Most notably, we could also add a variance term, and aim to minimize not only the mean, but also the variance.
We note that since our representations exist in a high dimensional space, we are comparing two covariance matrices in this case.
Nonetheless, a discrepancy term can be added to encourage these to be similar.
In our experiments, though, we found that adding such a variance term did not alter the results in any significant way.

As a remark, we note that even if two distributions share the same mean and covariance matrix, they can still be quite different.
However, if the distributions are close to Gaussian, then they would be characterized by their first two moments.
Of course, we cannot prove that the neural network will find representations that map the inputs to a Gaussian subspace of embedding space.
However, we did find experimentally that projecting the distributions of representations to random directions yielded one-dimensional distributions that looked Gaussian.
Thus, it is reasonable to believe that with a large enough training set, the representations would converge to something close to Gaussian.


\section{Experiments}
\label{sec:exp}

In this section, we begin with a description of the datasets we use as input in our experiments. 
Then, we evaluate each of our proposed techniques and baselines using a typical evaluation metric, which we will define. 
Finally, we discuss the sensitivity of each technique to changes in hyperparameters, as robustness to hyperparameter tuning hastens the training process.

\subsection{Datasets}

Due to the private and sensitive nature of the data, there is no publicly available large-scale enterprise email search dataset. 
As a result, the data we use comes from the search click logs of Gmail, a commercial email service. 
For our source dataset, we use click data from the entire Gmail search log. 
We then study domain adaptation to data arising from logs of four individual enterprise domains. 
The search logs for the source data consist of hundreds of millions of queries, whereas the target domains are significantly smaller. 
We chose the four largest domains (selected based on the total number of examples) as our target datasets and label them as A, B, C, and D. 
Domain A has around 65,000 total examples, B and C have around 50,000 each, and D has 30,000 queries, all significantly smaller than the source dataset.
The domains, including the source domain, are split into training and evaluation sets at approximately a $5:1$ ratio, and in such a way that all queries in the evaluation sets are performed on days after those in the training sets.
Each example consists of a query with six candidate documents, one of which is clicked. 

The goal of the model is to rank the six documents in such a way as to increase the likelihood of a higher ranked document being clicked.
In this way, clicks are regarded as the ground truth from which our model learns.

\subsection{Model Evaluation}

Our neural network models are implemented in TF-Ranking~\cite{TensorflowRanking2018}, a scalable open-source learning-to-rank Tensorflow~\cite{abadi16} library.
Our baselines are optimized over a number of hyperparameters, including learning rate, number of hidden layers, dimensions of layers, and batch normalization.
Specifically, we use a learning rate of $0.1$ and three hidden layers with dimensions 256, 128, and 64.
For each training procedure that takes as input both source and target training data, we also tried a number of different ratios of source to target training data.
However, we found that none of the procedures were sensitive to this ratio, as long as it was larger than $4:1$ or $5:1$.
Anything lower would cause overfitting, since the target dataset was so much smaller than the source.
Then, for the mean discrepancy (Eq.~\ref{eq:mmdloss}) and gradient reversal (Eq~\ref{eq:grloss}) losses, the $\lambda$ multipliers are optimized over $[0.3, 0.7, 1.0, 3.0, 7.0]$.

Model performance is evaluated using weighted mean reciprocal rank (WMRR), as proposed in ~\cite{wang16}. 
The weighted MRR (WMRR) is calculated using the one clicked document of any query as:
\begin{equation}
    \textrm{WMRR} = \frac{1}{\sum_1^{|E|} w_i} \sum_{i=1}^{|E|} w_i\frac{1}{\textrm{rank}_i}, \label{eq:mrr}
\end{equation}
where $E$ denotes the evaluation set, $\textrm{rank}_i$ denotes the position or rank of the clicked document for the $i$-th query, and $w_i$ denotes the bias correction weights.
The $w_i$ are inversely proportional to the probability of observing a click at the clicked position and are set using result randomization, as described in~\cite{wang16}. 
In addition to reporting the WMRR attained by each model, we also conduct statistical significance tests using the two-tailed paired t-test with $99\%$ confidence.

\subsection{Results and Discussion}

In this section, we provide our main results, as well as accompanying discussion.
We first describe the two standard baseline training methods.

\textbf{Standard Baselines:}
\begin{itemize}
    \item \textbf{Train on all}. This is the simplest baseline. We train a model on the entire source dataset, and then evaluate it on the specific domain test data.
    \item \textbf{Train on domain}. We form a training set consisting of only domain-specific data. As expected, there is not enough data to train a neural network capable of generalizing to test data. Not only does the model have a low WMRR, but we also see severe overfitting.
\end{itemize}

Since our goal is a thorough analysis of possible ways to train a prediction model for a specific domain, we also tried additional baselines.
Both the additional baselines typically outperformed the standard baselines, and we suggest that they should be used for comparison in any domain adaptation study.

\textbf{Additional Baselines:}
\begin{itemize}
    \item \textbf{Re-train}. This is typically known as vanilla transfer learning. We first load up a \textbf{train on all} model. Then, we re-train this model on only domain data with a lower learning rate. Specifically, we reduce the learning rate by a factor of $10$.
    \item \textbf{Batch-balance}. This model is similar to the \textbf{train on all}, in that the training set consists of the entire source dataset. The difference is that we enforce a certain proportion of target domain data in each training batch. 
\end{itemize}

The raw WMRR model evaluations from Equation~\ref{eq:mrr} for the baselines are provided in Table~\ref{tab:base}.

\begin{table}[!ht]
\caption{
WMRR evaluation results for all four baseline methods trained for each of the four separate domains, labeled A, B, C, and D.
}
\label{tab:base}
\begin{center}
{\renewcommand{\arraystretch}{1.4}
\begin{tabular}{c|c|c|c|c}
  & A & B & C & D \\
\hline    
   Train on all & 0.659 & 0.692 & 0.611 & 0.598 \\
   \hline    
   Train on domain & 0.639 & 0.694 & 0.573 & 0.579 \\
   \hline
   Re-train & 0.675 & 0.713 & \textbf{0.608} & 0.618 \\
   \hline
   Batch-balance & \textbf{0.682} & \textbf{0.715} & \textbf{0.608} & \textbf{0.621} \\
   \hline
\end{tabular}}
\end{center}
\end{table}

From Table~\ref{tab:base}, we can see that generally, \textbf{train on domain} performs the worst of all baselines. 
As noted before, this makes sense due to the fact that the domains do not provide enough training data for neural network models. 
Then, we have our standard training method \textbf{train on all}. 
Finally, \textbf{re-train} and \textbf{batch-balance} have roughly the same performance, with the latter performing slightly better for some domains. 
These are exactly the results we would expect, since these two methods are more involved than only training on domain-specific data or all the source data we have. 

Now, we briefly describe our proposed training methods. While they are described in full detail in Section~\ref{sec:methods}, we review them here for convenience.

\textbf{Domain Adaptation Methods:}
\begin{itemize}
    \item \textbf{Gradient Reversal}. As in the \textbf{Batch-balance} baseline, we enforce a certain proportion of domain data in each training batch. A discriminator is then added to try and distinguish, from the embeddings, whether data is from the specific domain. Gradient reversal is used on the embedding weights to fool the discriminator. The resulting loss is described in Equation \ref{eq:grloss}.
    \item \textbf{Mean Discrepancy}. Again, we enforce a certain proportion of domain data in each training batch. A regularization term is added to the standard cross entropy loss consisting of the difference in means of the source and target input embeddings. The resulting loss is described in Equation \ref{eq:mmdloss}.
\end{itemize}

Our main table of results is provided in Table~\ref{tab:core}.
Our numbers are recorded as relative improvement of our proposed domain adaptation methods to the corresponding baseline methods.

\begin{table*}[htp]
\caption{
WMRR evaluation results for adapting to four domains. Methods on the x-axis are compared to baselines on the y-axis, and results are given as relative improvement. An asterisk (*) denotes statistical significance relative to the baseline, according to our two-tailed paired t-test with $99\%$ confidence. Additionally, we underline the numbers where mean discrepancy is statistically significant over gradient reversal.
}
\label{tab:core}
\begin{center}
{\renewcommand{\arraystretch}{1.4}
\begin{tabular}{c|c|c||c|c|c}
  \textbf{Domain A} & \textbf{Gradient Reversal} & \textbf{Mean Discrepancy} & \textbf{Domain B} & \textbf{Gradient Reversal} & \textbf{Mean Discrepancy} \\
\hline    
   \textbf{Train on all} & +4.16\%* & \underline{\color{blue}+4.38\%*} & \textbf{Train on all} & +2.99\%* & \underline{\color{blue}+4.00\%*} \\
   \hline    
   \textbf{Train on domain} & +7.30\%* & \underline{\color{blue}+7.52\%*} & \textbf{Train on domain} & +2.77\%* & \underline{\color{blue}+3.79\%*} \\
   \hline    
   \textbf{Re-train} & +1.60\%* & \underline{\color{blue}+1.82\%*} & \textbf{Re-train} & +0.02\% & \underline{\color{blue}+1.00\%*} \\
   \hline    
   \textbf{Batch-balance} & +0.44\%* & \underline{\color{blue}+0.65\%*} & \textbf{Batch-balance} & -0.24\%* & \underline{\color{blue}+0.75\%*} \\
   \hline
   \hline
  \textbf{Domain C} & \textbf{Gradient Reversal} & \textbf{Mean Discrepancy} & \textbf{Domain D} & \textbf{Gradient Reversal} & \textbf{Mean Discrepancy} \\
\hline    
   \textbf{Train on all} & +1.37\%* & \underline{\color{blue}+1.49\%*} & \textbf{Train on all} & +2.45\%* & \underline{\color{blue}+5.30\%*}\\
   \hline    
   \textbf{Train on domain} & +8.04\%* & \underline{\color{blue}+8.17\%*} & \textbf{Train on domain} & +5.77\%* & \underline{\color{blue}+8.71\%*}\\
   \hline    
   \textbf{Re-train} & +1.85\%* & \underline{\color{blue}+1.97\%*} & \textbf{Re-train} & -0.92\%* & \underline{\color{blue}+1.83\%*}\\
   \hline    
   \textbf{Batch-balance} & +1.78\%* & \underline{\color{blue}+1.91\%*} & \textbf{Batch-balance} & -1.44\%* & \underline{\color{blue}+1.30\%*}\\
\end{tabular}}
\end{center}

\end{table*}

While the improvement changes from domain to domain, mean discrepancy is consistently the best performing proposed method.
In every instance, mean discrepancy achieves a higher WMRR than any of the baselines.
To give perspective, we note that improvements of $1\%$ are considered to be highly significant for our enterprise email search system.
Compared to \textbf{Re-train}, mean discrepancy achieves at least $1\%$ achievement, although for Domains A and B, it does not quite outperform batch-balance by this much.

While not directly listed as a comparison in the table, we also note that mean discrepancy outperforms gradient reversal in a statistically significant way.
In image classification, adversarial methods have shown to be better than those using maximum mean discrepancy~\cite{ghifary16, ganin14, tzeng15} for domain adaptation to an unsupervised domain.
But in our experiments, the opposite was shown to be true.
While we cannot say for certain why this difference exists, we provide a few possible explanations.
We first recall that the set of training examples aims to approximate the true distribution of inputs.
Since maximum mean discrepancy aims to minimize the difference between source and target distribution means, the method is highly dependent on how well the mean is approximated by training examples.
Since the typical size of datasets considered in prior work is quite small, often even smaller than a single target domain, the mean of the training example embeddings may not accurately represent the true mean.
With a thousand times more inputs in our source dataset, the mean is more accurately approximated and more useful for maximum mean discrepancy.

But even with good approximation of the true mean, it is fairly common to find situations in which reducing the MMD would not suffice to make two distributions indistinguishable.
One can easily imagine two distributions with identical mean that are still drastically different.
But as we showed in Table~\ref{tab:norms}, the source and target data embeddings in our problem do have significantly different means. 
Because of this, using MMD with the way we embed the query-document inputs ends up working very well for our domain adaptation task.

\subsection{Sensitivity Analysis}

In this section, we discuss the sensitivity of the mean discrepancy and gradient reversal methods.
In a regime with complicated inputs and long training times, a desired quality of any training method is robustness to hyperparameters.
Therefore, we compare the sensitivity of our two proposed training methods, maximum mean discrepancy and gradient reversal, to the changes in their respective parameters.
Recall that the set of parameters $\{\lambda_{D}, \lambda_{adv},  \lambda_{MMD}\}$ dictates the interactions of the distribution-matching correction terms -- part (3) of Figure~\ref{fig:diagram} -- with the ranking model's softmax cross-entropy term $\mathcal{L}_P$ (see Equations \ref{eq:grloss} and \ref{eq:mmdloss}).

Careful parameter tuning is especially important when using methods based on two-player games, such as gradient reversal.
In general, two-player games are sensitive to changes in parameters, making a good equilibrium difficult for gradient descent to find~\cite{roth17}.
As a result, we hypothesize that maximum mean discrepancy is the better training method for our enterprise email search setting, not only because of better overall WMRR, but also due to its robustness to changes in $\lambda_{MMD}$. In what follows, we provide some empirical evidence to back up this hypothesis.

\begin{figure}[!htp]
\begin{center}
\includegraphics[width=0.4\textwidth]{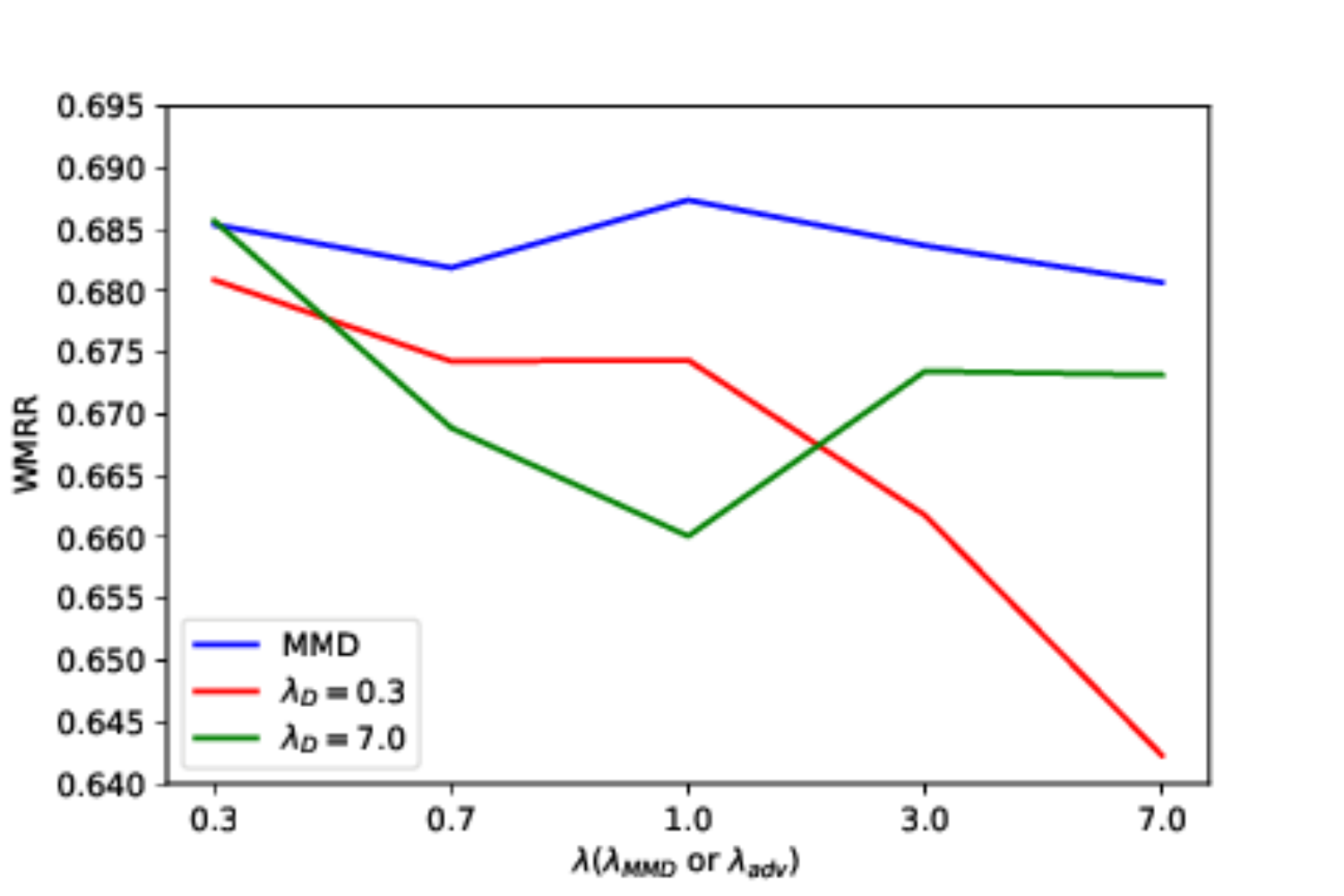}
\end{center}
\caption{WMRR curves on domain $A$ for both the mean discrepancy and the gradient reversal methods. The WMRR is plotted as a function of $\lambda_{MMD}$ for mean discrepancy and of $\lambda_{adv}$ for two fixed values of $\lambda_D$ for gradient reversal.}
\label{fig:rob}
\end{figure}

\begin{figure}[!htp]
\begin{center}
\includegraphics[width=0.45\textwidth]{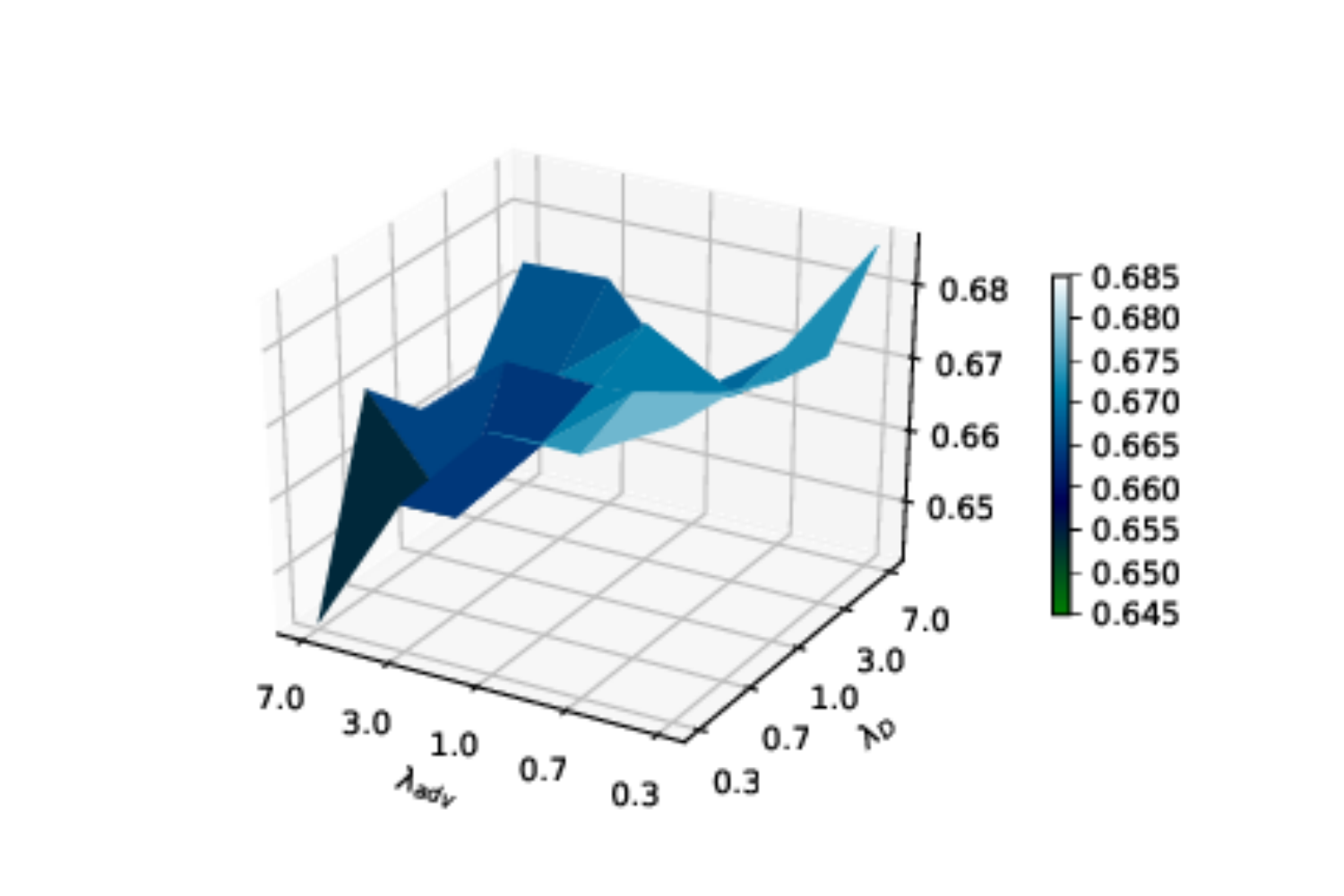}
\end{center}
\caption{WMRR values on domain $A$ when varying both $\lambda_{adv}$ and $\lambda_D$ parameters.}
\label{fig:gr_heat}
\end{figure}

First, in Figure~\ref{fig:rob}, we plot the resulting WMRR from training using a range of $\lambda$ values.
For the mean discrepancy method, this corresponds to $\lambda_{MMD}$, and for gradient reversal, we fix two values of $\lambda_{D}$ and vary $\lambda_{adv}$ on the x-axis.
The WMRR curves for fixed values of $\lambda_{D}$ are shown in order to give a direct comparison between the mean discrepancy and gradient reversal methods.
Additionally, we plot the entire three-dimensional surface of the WMRR metric as a function of both $\lambda_D$ and $\lambda_{adv}$ in Figure~\ref{fig:gr_heat}.

From Figure~\ref{fig:rob}, we can see that for a large range of $\lambda_{MMD}$ values, the resulting WMRR is relatively stable.
Specifically, in the range of $\lambda$ values we consider, from $0.3$ to $7.0$, the WMRR values remain within a $\pm 0.005$ range. Qualitatively, this robustness to tuning $\lambda_{MMD}$ holds for all domains.

On the other hand, the plots for gradient reversal support our conjecture that it is not as robust as mean discrepancy.
From Figure~\ref{fig:rob}, we see a direct comparison showing the greater sensitivity gradient reversal has relative to mean discrepancy.
Moreover, from Figure~\ref{fig:gr_heat}, we observe that the WMRR values are as far apart as $0.05$ from each other, nearly ten times as big a difference as we see for the mean discrepancy method. Again, similar trends could be observed for other domains.

\section{Conclusions}
\label{sec:conc}
In this paper, we studied the application of domain adaptation to learning-to-rank, specifically in the setting of enterprise email search. 
We developed a formal framework for integrating two techniques from the image classification literature for transfer learning, maximum mean discrepancy (MMD) and gradient reversal, in the learning-to-rank models. Our models were implemented in Tensorflow using a deep neural network that efficiently handles various features from query-document inputs, and provides embeddings to which the domain adaptation methods are applied.

The results from our experiments on a large-scale enterprise email search engine indicate that neither a single global model, nor simple transfer learning baselines are sufficient for achieving the optimal performance for individual domains. Overall, we show that maximum mean discrepancy (MMD) is the best technique for adapting the global learning-to-rank model to a small target domain in enterprise email search. MMD is not only the most effective method on all the tested domains, it also displays robustness to parameter changes.

One possible future direction of study is using regularization terms involving statistics other than the mean.
Another is to find more stable equilibrium for disciminator-based methods.
While much work has been done~\cite{salimans16, gulrajani17, arjovsky17, roth17} to improve the stability of models involving two-player games, thse are often very task-specific.
Finally, while in this paper we explored domain adaptation specifically in the enterprise email search setting, the proposed methods can easily generalize to other information retrieval tasks. 


\bibliographystyle{ACM-Reference-Format}
\bibliography{refs}


\begin{thebibliography}{46}


\ifx \showCODEN    \undefined \def \showCODEN     #1{\unskip}     \fi
\ifx \showDOI      \undefined \def \showDOI       #1{#1}\fi
\ifx \showISBNx    \undefined \def \showISBNx     #1{\unskip}     \fi
\ifx \showISBNxiii \undefined \def \showISBNxiii  #1{\unskip}     \fi
\ifx \showISSN     \undefined \def \showISSN      #1{\unskip}     \fi
\ifx \showLCCN     \undefined \def \showLCCN      #1{\unskip}     \fi
\ifx \shownote     \undefined \def \shownote      #1{#1}          \fi
\ifx \showarticletitle \undefined \def \showarticletitle #1{#1}   \fi
\ifx \showURL      \undefined \def \showURL       {\relax}        \fi
\providecommand\bibfield[2]{#2}
\providecommand\bibinfo[2]{#2}
\providecommand\natexlab[1]{#1}
\providecommand\showeprint[2][]{arXiv:#2}

\bibitem[\protect\citeauthoryear{Abadi, Barham, Chen, Chen, Davis, Dean, Devin,
  Ghemawat, Irving, Isard, et~al\mbox{.}}{Abadi et~al\mbox{.}}{2016}]%
        {abadi16}
\bibfield{author}{\bibinfo{person}{Mart{\'\i}n Abadi}, \bibinfo{person}{Paul
  Barham}, \bibinfo{person}{Jianmin Chen}, \bibinfo{person}{Zhifeng Chen},
  \bibinfo{person}{Andy Davis}, \bibinfo{person}{Jeffrey Dean},
  \bibinfo{person}{Matthieu Devin}, \bibinfo{person}{Sanjay Ghemawat},
  \bibinfo{person}{Geoffrey Irving}, \bibinfo{person}{Michael Isard},
  {et~al\mbox{.}}} \bibinfo{year}{2016}\natexlab{}.
\newblock \showarticletitle{Tensorflow: a system for large-scale machine
  learning.}. In \bibinfo{booktitle}{\emph{Operating Systems Design and
  Implementation}}, Vol.~\bibinfo{volume}{16}. \bibinfo{pages}{265--283}.
\newblock


\bibitem[\protect\citeauthoryear{Ai, Dumais, Craswell, and Liebling}{Ai
  et~al\mbox{.}}{2017}]%
        {ai17}
\bibfield{author}{\bibinfo{person}{Qingyao Ai}, \bibinfo{person}{Susan~T
  Dumais}, \bibinfo{person}{Nick Craswell}, {and} \bibinfo{person}{Dan
  Liebling}.} \bibinfo{year}{2017}\natexlab{}.
\newblock \showarticletitle{Characterizing email search using large-scale
  behavioral logs and surveys}. In \bibinfo{booktitle}{\emph{Proceedings of the
  26th International Conference on World Wide Web}}.
  \bibinfo{pages}{1511--1520}.
\newblock


\bibitem[\protect\citeauthoryear{Arjovsky and Bottou}{Arjovsky and
  Bottou}{2017}]%
        {arjovsky17}
\bibfield{author}{\bibinfo{person}{Martin Arjovsky} {and}
  \bibinfo{person}{L{\'e}on Bottou}.} \bibinfo{year}{2017}\natexlab{}.
\newblock \showarticletitle{Towards principled methods for training generative
  adversarial networks}.
\newblock \bibinfo{journal}{\emph{arXiv preprint arXiv:1701.04862}}
  (\bibinfo{year}{2017}).
\newblock


\bibitem[\protect\citeauthoryear{Borisov, Markov, de~Rijke, and
  Serdyukov}{Borisov et~al\mbox{.}}{2016}]%
        {borisov16}
\bibfield{author}{\bibinfo{person}{Alexey Borisov}, \bibinfo{person}{Ilya
  Markov}, \bibinfo{person}{Maarten de Rijke}, {and} \bibinfo{person}{Pavel
  Serdyukov}.} \bibinfo{year}{2016}\natexlab{}.
\newblock \showarticletitle{A neural click model for web search}. In
  \bibinfo{booktitle}{\emph{Proceedings of the 25th International Conference on
  World Wide Web}}. \bibinfo{pages}{531--541}.
\newblock


\bibitem[\protect\citeauthoryear{Burges, Shaked, Renshaw, Lazier, Deeds,
  Hamilton, and Hullender}{Burges et~al\mbox{.}}{2005}]%
        {burges05}
\bibfield{author}{\bibinfo{person}{Chris Burges}, \bibinfo{person}{Tal Shaked},
  \bibinfo{person}{Erin Renshaw}, \bibinfo{person}{Ari Lazier},
  \bibinfo{person}{Matt Deeds}, \bibinfo{person}{Nicole Hamilton}, {and}
  \bibinfo{person}{Greg Hullender}.} \bibinfo{year}{2005}\natexlab{}.
\newblock \showarticletitle{Learning to rank using gradient descent}. In
  \bibinfo{booktitle}{\emph{Proceedings of the 22nd International Conference on
  Machine Learning}}. \bibinfo{pages}{89--96}.
\newblock


\bibitem[\protect\citeauthoryear{Burges}{Burges}{2010}]%
        {burges10}
\bibfield{author}{\bibinfo{person}{Christopher~JC Burges}.}
  \bibinfo{year}{2010}\natexlab{}.
\newblock \showarticletitle{From Ranknet to Lambdarank to Lambdamart: An
  overview}.
\newblock \bibinfo{journal}{\emph{Learning}} \bibinfo{volume}{11},
  \bibinfo{number}{23-581} (\bibinfo{year}{2010}), \bibinfo{pages}{81}.
\newblock


\bibitem[\protect\citeauthoryear{Cao, Qin, Liu, Tsai, and Li}{Cao
  et~al\mbox{.}}{2007}]%
        {cao07}
\bibfield{author}{\bibinfo{person}{Zhe Cao}, \bibinfo{person}{Tao Qin},
  \bibinfo{person}{Tie-Yan Liu}, \bibinfo{person}{Ming-Feng Tsai}, {and}
  \bibinfo{person}{Hang Li}.} \bibinfo{year}{2007}\natexlab{}.
\newblock \showarticletitle{Learning to rank: from pairwise approach to
  listwise approach}. In \bibinfo{booktitle}{\emph{Proceedings of the 24th
  International Conference on Machine Learning}}. \bibinfo{pages}{129--136}.
\newblock


\bibitem[\protect\citeauthoryear{Carmel, Halawi, Lewin-Eytan, Maarek, and
  Raviv}{Carmel et~al\mbox{.}}{2015}]%
        {carmel15}
\bibfield{author}{\bibinfo{person}{David Carmel}, \bibinfo{person}{Guy Halawi},
  \bibinfo{person}{Liane Lewin-Eytan}, \bibinfo{person}{Yoelle Maarek}, {and}
  \bibinfo{person}{Ariel Raviv}.} \bibinfo{year}{2015}\natexlab{}.
\newblock \showarticletitle{Rank by time or by relevance?: Revisiting email
  search}. In \bibinfo{booktitle}{\emph{Proceedings of the 24th ACM
  International on Conference on Information and Knowledge Management}}.
  \bibinfo{pages}{283--292}.
\newblock


\bibitem[\protect\citeauthoryear{Cohen, Mitra, Hofmann, and Croft}{Cohen
  et~al\mbox{.}}{2018}]%
        {cohen18}
\bibfield{author}{\bibinfo{person}{Daniel Cohen}, \bibinfo{person}{Bhaskar
  Mitra}, \bibinfo{person}{Katja Hofmann}, {and} \bibinfo{person}{W~Bruce
  Croft}.} \bibinfo{year}{2018}\natexlab{}.
\newblock \showarticletitle{Cross domain regularization for neural ranking
  models using adversarial learning}.
\newblock \bibinfo{journal}{\emph{arXiv preprint arXiv:1805.03403}}
  (\bibinfo{year}{2018}).
\newblock


\bibitem[\protect\citeauthoryear{Dehghani, Zamani, Severyn, Kamps, and
  Croft}{Dehghani et~al\mbox{.}}{2017}]%
        {dehghani17}
\bibfield{author}{\bibinfo{person}{Mostafa Dehghani}, \bibinfo{person}{Hamed
  Zamani}, \bibinfo{person}{Aliaksei Severyn}, \bibinfo{person}{Jaap Kamps},
  {and} \bibinfo{person}{W~Bruce Croft}.} \bibinfo{year}{2017}\natexlab{}.
\newblock \showarticletitle{Neural ranking models with weak supervision}. In
  \bibinfo{booktitle}{\emph{Proceedings of the 40th International ACM SIGIR
  Conference on Research and Development in Information Retrieval}}.
  \bibinfo{pages}{65--74}.
\newblock


\bibitem[\protect\citeauthoryear{Diakonikolas, Kamath, Kane, Li, Moitra, and
  Stewart}{Diakonikolas et~al\mbox{.}}{2016}]%
        {diakonikolas16}
\bibfield{author}{\bibinfo{person}{Ilias Diakonikolas}, \bibinfo{person}{Gautam
  Kamath}, \bibinfo{person}{Daniel~M Kane}, \bibinfo{person}{Jerry Li},
  \bibinfo{person}{Ankur Moitra}, {and} \bibinfo{person}{Alistair Stewart}.}
  \bibinfo{year}{2016}\natexlab{}.
\newblock \showarticletitle{Robust estimators in high dimensions without the
  computational intractability}. In \bibinfo{booktitle}{\emph{IEEE 57th Annual
  Symposium on Foundations of Computer Science}}. \bibinfo{pages}{655--664}.
\newblock


\bibitem[\protect\citeauthoryear{Dumais, Cutrell, Cadiz, Jancke, Sarin, and
  Robbins}{Dumais et~al\mbox{.}}{2016}]%
        {dumais16}
\bibfield{author}{\bibinfo{person}{Susan Dumais}, \bibinfo{person}{Edward
  Cutrell}, \bibinfo{person}{Jonathan~J Cadiz}, \bibinfo{person}{Gavin Jancke},
  \bibinfo{person}{Raman Sarin}, {and} \bibinfo{person}{Daniel~C Robbins}.}
  \bibinfo{year}{2016}\natexlab{}.
\newblock \showarticletitle{Stuff I've seen: a system for personal information
  retrieval and re-use}. In \bibinfo{booktitle}{\emph{ACM SIGIR Forum}},
  Vol.~\bibinfo{volume}{49}. \bibinfo{pages}{28--35}.
\newblock


\bibitem[\protect\citeauthoryear{Edizel, Mantrach, and Bai}{Edizel
  et~al\mbox{.}}{2017}]%
        {edizel17}
\bibfield{author}{\bibinfo{person}{Bora Edizel}, \bibinfo{person}{Amin
  Mantrach}, {and} \bibinfo{person}{Xiao Bai}.}
  \bibinfo{year}{2017}\natexlab{}.
\newblock \showarticletitle{Deep Character-Level Click-Through Rate Prediction
  for Sponsored Search}.
\newblock \bibinfo{journal}{\emph{arXiv preprint arXiv:1707.02158}}
  (\bibinfo{year}{2017}).
\newblock


\bibitem[\protect\citeauthoryear{Friedman}{Friedman}{2001}]%
        {friedman01}
\bibfield{author}{\bibinfo{person}{Jerome~H Friedman}.}
  \bibinfo{year}{2001}\natexlab{}.
\newblock \showarticletitle{Greedy function approximation: a gradient boosting
  machine}.
\newblock \bibinfo{journal}{\emph{Annals of Statistics}}
  (\bibinfo{year}{2001}), \bibinfo{pages}{1189--1232}.
\newblock


\bibitem[\protect\citeauthoryear{Ganin and Lempitsky}{Ganin and
  Lempitsky}{2014}]%
        {ganin14}
\bibfield{author}{\bibinfo{person}{Yaroslav Ganin} {and}
  \bibinfo{person}{Victor Lempitsky}.} \bibinfo{year}{2014}\natexlab{}.
\newblock \showarticletitle{Unsupervised domain adaptation by backpropagation}.
\newblock \bibinfo{journal}{\emph{arXiv preprint arXiv:1409.7495}}
  (\bibinfo{year}{2014}).
\newblock


\bibitem[\protect\citeauthoryear{Ganin, Ustinova, Ajakan, Germain, Larochelle,
  Laviolette, Marchand, and Lempitsky}{Ganin et~al\mbox{.}}{2016}]%
        {ganin16}
\bibfield{author}{\bibinfo{person}{Yaroslav Ganin}, \bibinfo{person}{Evgeniya
  Ustinova}, \bibinfo{person}{Hana Ajakan}, \bibinfo{person}{Pascal Germain},
  \bibinfo{person}{Hugo Larochelle}, \bibinfo{person}{Fran{\c{c}}ois
  Laviolette}, \bibinfo{person}{Mario Marchand}, {and} \bibinfo{person}{Victor
  Lempitsky}.} \bibinfo{year}{2016}\natexlab{}.
\newblock \showarticletitle{Domain-adversarial training of neural networks}.
\newblock \bibinfo{journal}{\emph{The Journal of Machine Learning Research}}
  \bibinfo{volume}{17}, \bibinfo{number}{1} (\bibinfo{year}{2016}),
  \bibinfo{pages}{2096--2030}.
\newblock


\bibitem[\protect\citeauthoryear{Ghifary, Kleijn, Zhang, Balduzzi, and
  Li}{Ghifary et~al\mbox{.}}{2016}]%
        {ghifary16}
\bibfield{author}{\bibinfo{person}{Muhammad Ghifary},
  \bibinfo{person}{W~Bastiaan Kleijn}, \bibinfo{person}{Mengjie Zhang},
  \bibinfo{person}{David Balduzzi}, {and} \bibinfo{person}{Wen Li}.}
  \bibinfo{year}{2016}\natexlab{}.
\newblock \showarticletitle{Deep reconstruction-classification networks for
  unsupervised domain adaptation}. In \bibinfo{booktitle}{\emph{European
  Conference on Computer Vision}}. \bibinfo{pages}{597--613}.
\newblock


\bibitem[\protect\citeauthoryear{Goodfellow, Pouget-Abadie, Mirza, Xu,
  Warde-Farley, Ozair, Courville, and Bengio}{Goodfellow et~al\mbox{.}}{2014}]%
        {goodfellow14}
\bibfield{author}{\bibinfo{person}{Ian Goodfellow}, \bibinfo{person}{Jean
  Pouget-Abadie}, \bibinfo{person}{Mehdi Mirza}, \bibinfo{person}{Bing Xu},
  \bibinfo{person}{David Warde-Farley}, \bibinfo{person}{Sherjil Ozair},
  \bibinfo{person}{Aaron Courville}, {and} \bibinfo{person}{Yoshua Bengio}.}
  \bibinfo{year}{2014}\natexlab{}.
\newblock \showarticletitle{Generative adversarial nets}. In
  \bibinfo{booktitle}{\emph{Advances in Neural Information Processing
  Systems}}. \bibinfo{pages}{2672--2680}.
\newblock


\bibitem[\protect\citeauthoryear{Gretton, Smola, Huang, Schmittfull, Borgwardt,
  and Sch{\"o}lkopf}{Gretton et~al\mbox{.}}{2009}]%
        {gretton09}
\bibfield{author}{\bibinfo{person}{A. Gretton}, \bibinfo{person}{AJ. Smola},
  \bibinfo{person}{J. Huang}, \bibinfo{person}{M. Schmittfull},
  \bibinfo{person}{KM. Borgwardt}, {and} \bibinfo{person}{B. Sch{\"o}lkopf}.}
  \bibinfo{year}{2009}\natexlab{}.
\newblock \bibinfo{booktitle}{\emph{Covariate shift and local learning by
  distribution matching}}.
\newblock \bibinfo{publisher}{MIT Press}, \bibinfo{address}{Cambridge, MA,
  USA}, \bibinfo{pages}{131--160}.
\newblock


\bibitem[\protect\citeauthoryear{Grevet, Choi, Kumar, and Gilbert}{Grevet
  et~al\mbox{.}}{2014}]%
        {grevet14}
\bibfield{author}{\bibinfo{person}{Catherine Grevet}, \bibinfo{person}{David
  Choi}, \bibinfo{person}{Debra Kumar}, {and} \bibinfo{person}{Eric Gilbert}.}
  \bibinfo{year}{2014}\natexlab{}.
\newblock \showarticletitle{Overload is overloaded: Email in the age of Gmail}.
  In \bibinfo{booktitle}{\emph{Proceedings of the Sigchi Conference on Human
  Factors in Computing Systems}}. \bibinfo{pages}{793--802}.
\newblock


\bibitem[\protect\citeauthoryear{Gulrajani, Ahmed, Arjovsky, Dumoulin, and
  Courville}{Gulrajani et~al\mbox{.}}{2017}]%
        {gulrajani17}
\bibfield{author}{\bibinfo{person}{Ishaan Gulrajani}, \bibinfo{person}{Faruk
  Ahmed}, \bibinfo{person}{Martin Arjovsky}, \bibinfo{person}{Vincent
  Dumoulin}, {and} \bibinfo{person}{Aaron~C Courville}.}
  \bibinfo{year}{2017}\natexlab{}.
\newblock \showarticletitle{Improved training of Wasserstein Gans}. In
  \bibinfo{booktitle}{\emph{Advances in Neural Information Processing
  Systems}}. \bibinfo{pages}{5767--5777}.
\newblock


\bibitem[\protect\citeauthoryear{Guo, Fan, Ai, and Croft}{Guo
  et~al\mbox{.}}{2016}]%
        {guo16}
\bibfield{author}{\bibinfo{person}{Jiafeng Guo}, \bibinfo{person}{Yixing Fan},
  \bibinfo{person}{Qingyao Ai}, {and} \bibinfo{person}{W~Bruce Croft}.}
  \bibinfo{year}{2016}\natexlab{}.
\newblock \showarticletitle{A deep relevance matching model for ad-hoc
  retrieval}. In \bibinfo{booktitle}{\emph{Proceedings of the 25th ACM
  International on Conference on Information and Knowledge Management}}.
  \bibinfo{pages}{55--64}.
\newblock


\bibitem[\protect\citeauthoryear{Hawking}{Hawking}{2010}]%
        {Hawking:2010}
\bibfield{author}{\bibinfo{person}{David Hawking}.}
  \bibinfo{year}{2010}\natexlab{}.
\newblock \showarticletitle{Enterprise Search}.
\newblock In \bibinfo{booktitle}{\emph{Modern Information Retrieval, 2nd
  Edition}}, \bibfield{editor}{\bibinfo{person}{Ricardo Baeza-Yates} {and}
  \bibinfo{person}{Berthier Ribeiro-Neto}} (Eds.).
  \bibinfo{publisher}{Addison-Wesley}, \bibinfo{pages}{645--686}.
\newblock
\urldef\tempurl%
\url{http://david-hawking.net/pubs/ModernIR2_Hawking_chapter.pdf}
\showURL{%
\tempurl}


\bibitem[\protect\citeauthoryear{Huang, He, Gao, Deng, Acero, and Heck}{Huang
  et~al\mbox{.}}{2013}]%
        {huang13}
\bibfield{author}{\bibinfo{person}{Po-Sen Huang}, \bibinfo{person}{Xiaodong
  He}, \bibinfo{person}{Jianfeng Gao}, \bibinfo{person}{Li Deng},
  \bibinfo{person}{Alex Acero}, {and} \bibinfo{person}{Larry Heck}.}
  \bibinfo{year}{2013}\natexlab{}.
\newblock \showarticletitle{Learning deep structured semantic models for web
  search using clickthrough data}. In \bibinfo{booktitle}{\emph{Proceedings of
  the 22nd ACM International Conference on Conference on Information \&
  Knowledge Management}}. \bibinfo{pages}{2333--2338}.
\newblock


\bibitem[\protect\citeauthoryear{Joachims}{Joachims}{2002}]%
        {joachims02}
\bibfield{author}{\bibinfo{person}{Thorsten Joachims}.}
  \bibinfo{year}{2002}\natexlab{}.
\newblock \showarticletitle{Optimizing search engines using clickthrough data}.
  In \bibinfo{booktitle}{\emph{Proceedings of the eighth ACM SIGKDD
  International Conference on Knowledge Discovery and Data Mining}}.
  \bibinfo{pages}{133--142}.
\newblock


\bibitem[\protect\citeauthoryear{Kim, Craswell, Dumais, Radlinski, and Liu}{Kim
  et~al\mbox{.}}{2017}]%
        {kim17}
\bibfield{author}{\bibinfo{person}{Jin~Young Kim}, \bibinfo{person}{Nick
  Craswell}, \bibinfo{person}{Susan Dumais}, \bibinfo{person}{Filip Radlinski},
  {and} \bibinfo{person}{Fang Liu}.} \bibinfo{year}{2017}\natexlab{}.
\newblock \showarticletitle{Understanding and Modeling Success in Email
  Search}. In \bibinfo{booktitle}{\emph{Proceedings of the 40th International
  ACM SIGIR Conference on Research and Development in Information Retrieval}}.
  \bibinfo{pages}{265--274}.
\newblock


\bibitem[\protect\citeauthoryear{Kruschwitz and Hull}{Kruschwitz and
  Hull}{2017}]%
        {Kruschwitz+Hull:2017}
\bibfield{author}{\bibinfo{person}{Udo Kruschwitz} {and}
  \bibinfo{person}{Charlie Hull}.} \bibinfo{year}{2017}\natexlab{}.
\newblock \showarticletitle{Searching the Enterprise}.
\newblock \bibinfo{journal}{\emph{Foundations and Trends® in Information
  Retrieval}} \bibinfo{volume}{11}, \bibinfo{number}{1} (\bibinfo{year}{2017}),
  \bibinfo{pages}{1--142}.
\newblock
\showISSN{1554-0669}
\urldef\tempurl%
\url{https://doi.org/10.1561/1500000053}
\showDOI{\tempurl}


\bibitem[\protect\citeauthoryear{Lai, Rao, and Vempala}{Lai
  et~al\mbox{.}}{2016}]%
        {lai16}
\bibfield{author}{\bibinfo{person}{Kevin~A Lai}, \bibinfo{person}{Anup~B Rao},
  {and} \bibinfo{person}{Santosh Vempala}.} \bibinfo{year}{2016}\natexlab{}.
\newblock \showarticletitle{Agnostic estimation of mean and covariance}. In
  \bibinfo{booktitle}{\emph{IEEE 57th Annual Symposium on Foundations of
  Computer Science}}. \bibinfo{pages}{665--674}.
\newblock


\bibitem[\protect\citeauthoryear{LeCun, Bengio, and Hinton}{LeCun
  et~al\mbox{.}}{2015}]%
        {lecun15}
\bibfield{author}{\bibinfo{person}{Yann LeCun}, \bibinfo{person}{Yoshua
  Bengio}, {and} \bibinfo{person}{Geoffrey Hinton}.}
  \bibinfo{year}{2015}\natexlab{}.
\newblock \showarticletitle{Deep Learning}.
\newblock \bibinfo{journal}{\emph{Nature}} \bibinfo{volume}{521},
  \bibinfo{number}{7553} (\bibinfo{year}{2015}), \bibinfo{pages}{436}.
\newblock


\bibitem[\protect\citeauthoryear{Liu and Tuzel}{Liu and Tuzel}{2016}]%
        {liu16}
\bibfield{author}{\bibinfo{person}{Ming-Yu Liu} {and} \bibinfo{person}{Oncel
  Tuzel}.} \bibinfo{year}{2016}\natexlab{}.
\newblock \showarticletitle{Coupled generative adversarial networks}. In
  \bibinfo{booktitle}{\emph{Advances in neural information processing
  systems}}. \bibinfo{pages}{469--477}.
\newblock


\bibitem[\protect\citeauthoryear{Long, Yao, Dai, Tian, Luo, and Mei}{Long
  et~al\mbox{.}}{2018}]%
        {long18}
\bibfield{author}{\bibinfo{person}{Fuchen Long}, \bibinfo{person}{Ting Yao},
  \bibinfo{person}{Qi Dai}, \bibinfo{person}{Xinmei Tian},
  \bibinfo{person}{Jiebo Luo}, {and} \bibinfo{person}{Tao Mei}.}
  \bibinfo{year}{2018}\natexlab{}.
\newblock \showarticletitle{Deep Domain Adaptation Hashing with Adversarial
  Learning}. In \bibinfo{booktitle}{\emph{the 41st International ACM SIGIR
  Conference on Research; Development in Information Retrieval}}.
\newblock


\bibitem[\protect\citeauthoryear{Long, Cao, Wang, and Jordan}{Long
  et~al\mbox{.}}{2015}]%
        {long15}
\bibfield{author}{\bibinfo{person}{Mingsheng Long}, \bibinfo{person}{Yue Cao},
  \bibinfo{person}{Jianmin Wang}, {and} \bibinfo{person}{Michael~I Jordan}.}
  \bibinfo{year}{2015}\natexlab{}.
\newblock \showarticletitle{Learning transferable features with deep adaptation
  networks}.
\newblock \bibinfo{journal}{\emph{arXiv preprint arXiv:1502.02791}}
  (\bibinfo{year}{2015}).
\newblock


\bibitem[\protect\citeauthoryear{Mao, Shen, and Chung}{Mao
  et~al\mbox{.}}{2018}]%
        {mao18}
\bibfield{author}{\bibinfo{person}{Sitong Mao}, \bibinfo{person}{Xiao Shen},
  {and} \bibinfo{person}{Fu-lai Chung}.} \bibinfo{year}{2018}\natexlab{}.
\newblock \showarticletitle{Deep Domain Adaptation Based on Multi-layer Joint
  Kernelized Distance}. In \bibinfo{booktitle}{\emph{the 41st International ACM
  SIGIR Conference on Research; Development in Information Retrieval}}.
\newblock


\bibitem[\protect\citeauthoryear{Mitra and Craswell}{Mitra and
  Craswell}{2017}]%
        {Mitra2017}
\bibfield{author}{\bibinfo{person}{Bhaskar Mitra} {and} \bibinfo{person}{Nick
  Craswell}.} \bibinfo{year}{2017}\natexlab{}.
\newblock \showarticletitle{Neural Models for Information Retrieval}.
\newblock \bibinfo{journal}{\emph{arXiv preprint arXiv:1805.03403}}
  (\bibinfo{year}{2017}).
\newblock


\bibitem[\protect\citeauthoryear{Mohri, Rostamizadeh, and Talwalkar}{Mohri
  et~al\mbox{.}}{2018}]%
        {mohri18}
\bibfield{author}{\bibinfo{person}{Mehryar Mohri}, \bibinfo{person}{Afshin
  Rostamizadeh}, {and} \bibinfo{person}{Ameet Talwalkar}.}
  \bibinfo{year}{2018}\natexlab{}.
\newblock \bibinfo{booktitle}{\emph{Foundations of machine learning}}.
\newblock \bibinfo{publisher}{MIT press}.
\newblock


\bibitem[\protect\citeauthoryear{Pasumarthi, Bruch, Wang, Li, Bendersky,
  Najork, Pfeifer, Golbandi, Anil, and Wolf}{Pasumarthi et~al\mbox{.}}{2019}]%
        {TensorflowRanking2018}
\bibfield{author}{\bibinfo{person}{Rama~Kumar Pasumarthi},
  \bibinfo{person}{Sebastian Bruch}, \bibinfo{person}{Xuanhui Wang},
  \bibinfo{person}{Cheng Li}, \bibinfo{person}{Michael Bendersky},
  \bibinfo{person}{Marc Najork}, \bibinfo{person}{Jan Pfeifer},
  \bibinfo{person}{Nadav Golbandi}, \bibinfo{person}{Rohan Anil}, {and}
  \bibinfo{person}{Stephan Wolf}.} \bibinfo{year}{2019}\natexlab{}.
\newblock \showarticletitle{TF-Ranking: Scalable TensorFlow Library for
  Learning-to-Rank}. In \bibinfo{booktitle}{\emph{Proceedings of the 25th ACM
  SIGKDD International Conference on Knowledge Discovery and Data Mining}}.
  \bibinfo{pages}{(to appear)}.
\newblock


\bibitem[\protect\citeauthoryear{Roth, Lucchi, Nowozin, and Hofmann}{Roth
  et~al\mbox{.}}{2017}]%
        {roth17}
\bibfield{author}{\bibinfo{person}{Kevin Roth}, \bibinfo{person}{Aurelien
  Lucchi}, \bibinfo{person}{Sebastian Nowozin}, {and} \bibinfo{person}{Thomas
  Hofmann}.} \bibinfo{year}{2017}\natexlab{}.
\newblock \showarticletitle{Stabilizing training of generative adversarial
  networks through regularization}. In \bibinfo{booktitle}{\emph{Advances in
  Neural Information Processing Systems}}. \bibinfo{pages}{2018--2028}.
\newblock


\bibitem[\protect\citeauthoryear{Salimans, Goodfellow, Zaremba, Cheung,
  Radford, and Chen}{Salimans et~al\mbox{.}}{2016}]%
        {salimans16}
\bibfield{author}{\bibinfo{person}{Tim Salimans}, \bibinfo{person}{Ian
  Goodfellow}, \bibinfo{person}{Wojciech Zaremba}, \bibinfo{person}{Vicki
  Cheung}, \bibinfo{person}{Alec Radford}, {and} \bibinfo{person}{Xi Chen}.}
  \bibinfo{year}{2016}\natexlab{}.
\newblock \showarticletitle{Improved techniques for training gans}. In
  \bibinfo{booktitle}{\emph{Advances in Neural Information Processing
  Systems}}. \bibinfo{pages}{2234--2242}.
\newblock


\bibitem[\protect\citeauthoryear{Shen, Karimzadehgan, Bendersky, Qin, and
  Metzler}{Shen et~al\mbox{.}}{2018}]%
        {meltzer18}
\bibfield{author}{\bibinfo{person}{Jiaming Shen}, \bibinfo{person}{Maryam
  Karimzadehgan}, \bibinfo{person}{Michael Bendersky}, \bibinfo{person}{Zhen
  Qin}, {and} \bibinfo{person}{Don Metzler}.} \bibinfo{year}{2018}\natexlab{}.
\newblock \showarticletitle{Multi-Task Learning for Personal Search Ranking
  with Query Clustering}. In \bibinfo{booktitle}{\emph{Proceedings of ACM
  Conference on Information and Knowledge Management}}.
\newblock


\bibitem[\protect\citeauthoryear{Sun and Saenko}{Sun and Saenko}{2016}]%
        {sun16}
\bibfield{author}{\bibinfo{person}{Baochen Sun} {and} \bibinfo{person}{Kate
  Saenko}.} \bibinfo{year}{2016}\natexlab{}.
\newblock \showarticletitle{Deep coral: Correlation alignment for deep domain
  adaptation}. In \bibinfo{booktitle}{\emph{European Conference on Computer
  Vision}}. \bibinfo{pages}{443--450}.
\newblock


\bibitem[\protect\citeauthoryear{Tzeng, Hoffman, Darrell, and Saenko}{Tzeng
  et~al\mbox{.}}{2015}]%
        {tzeng15}
\bibfield{author}{\bibinfo{person}{Eric Tzeng}, \bibinfo{person}{Judy Hoffman},
  \bibinfo{person}{Trevor Darrell}, {and} \bibinfo{person}{Kate Saenko}.}
  \bibinfo{year}{2015}\natexlab{}.
\newblock \showarticletitle{Simultaneous deep transfer across domains and
  tasks}. In \bibinfo{booktitle}{\emph{Proceedings of the IEEE International
  Conference on Computer Vision}}. \bibinfo{pages}{4068--4076}.
\newblock


\bibitem[\protect\citeauthoryear{Tzeng, Hoffman, Saenko, and Darrell}{Tzeng
  et~al\mbox{.}}{2017}]%
        {tzeng17}
\bibfield{author}{\bibinfo{person}{Eric Tzeng}, \bibinfo{person}{Judy Hoffman},
  \bibinfo{person}{Kate Saenko}, {and} \bibinfo{person}{Trevor Darrell}.}
  \bibinfo{year}{2017}\natexlab{}.
\newblock \showarticletitle{Adversarial discriminative domain adaptation}. In
  \bibinfo{booktitle}{\emph{Computer Vision and Pattern Recognition}}.
  \bibinfo{pages}{4}.
\newblock


\bibitem[\protect\citeauthoryear{Tzeng, Hoffman, Zhang, Saenko, and
  Darrell}{Tzeng et~al\mbox{.}}{2014}]%
        {tzeng14}
\bibfield{author}{\bibinfo{person}{Eric Tzeng}, \bibinfo{person}{Judy Hoffman},
  \bibinfo{person}{Ning Zhang}, \bibinfo{person}{Kate Saenko}, {and}
  \bibinfo{person}{Trevor Darrell}.} \bibinfo{year}{2014}\natexlab{}.
\newblock \showarticletitle{Deep domain confusion: Maximizing for domain
  invariance}.
\newblock \bibinfo{journal}{\emph{arXiv preprint arXiv:1412.3474}}
  (\bibinfo{year}{2014}).
\newblock


\bibitem[\protect\citeauthoryear{Wang, Bendersky, Metzler, and Najork}{Wang
  et~al\mbox{.}}{2016}]%
        {wang16}
\bibfield{author}{\bibinfo{person}{Xuanhui Wang}, \bibinfo{person}{Michael
  Bendersky}, \bibinfo{person}{Donald Metzler}, {and} \bibinfo{person}{Marc
  Najork}.} \bibinfo{year}{2016}\natexlab{}.
\newblock \showarticletitle{Learning to rank with selection bias in personal
  search}. In \bibinfo{booktitle}{\emph{Proceedings of the 39th International
  ACM SIGIR Conference on Research and Development in Information Retrieval}}.
  \bibinfo{pages}{115--124}.
\newblock


\bibitem[\protect\citeauthoryear{Xia, Liu, Wang, Zhang, and Li}{Xia
  et~al\mbox{.}}{2008}]%
        {xia08}
\bibfield{author}{\bibinfo{person}{Fen Xia}, \bibinfo{person}{Tie-Yan Liu},
  \bibinfo{person}{Jue Wang}, \bibinfo{person}{Wensheng Zhang}, {and}
  \bibinfo{person}{Hang Li}.} \bibinfo{year}{2008}\natexlab{}.
\newblock \showarticletitle{Listwise approach to learning to rank: theory and
  algorithm}. In \bibinfo{booktitle}{\emph{Proceedings of the 25th
  International Conference on Machine Learning}}. \bibinfo{pages}{1192--1199}.
\newblock


\bibitem[\protect\citeauthoryear{Zamani, Bendersky, Wang, and Zhang}{Zamani
  et~al\mbox{.}}{2017}]%
        {zamani17}
\bibfield{author}{\bibinfo{person}{Hamed Zamani}, \bibinfo{person}{Michael
  Bendersky}, \bibinfo{person}{Xuanhui Wang}, {and} \bibinfo{person}{Mingyang
  Zhang}.} \bibinfo{year}{2017}\natexlab{}.
\newblock \showarticletitle{Situational context for ranking in personal
  search}. In \bibinfo{booktitle}{\emph{Proceedings of the 26th International
  Conference on World Wide Web}}. \bibinfo{pages}{1531--1540}.
\newblock


\end{thebibliography}

\end{document}